\documentclass[superscriptaddress,aps,prb,reprint]{revtex4-2}

\usepackage{amsmath,amssymb}
\usepackage{graphicx}
\usepackage{bm}
\usepackage{multirow}

\usepackage{hyperref}
\hypersetup{colorlinks=true, linkcolor=blue, citecolor=blue, urlcolor=blue}

\usepackage{xcolor}

\begin{document}

\title{Assessing the classicality of photon echo from excitons \\ in lead halide perovskite nanocrystals}

\author{George Alkhalil}
\affiliation{Experimentelle Physik 2, Technische Universit\"at Dortmund, 44227 Dortmund, Germany}
\author{Hendrik Rose}
\affiliation{Institute for Photonic Quantum Systems (PhoQS), Paderborn University, D-33098 Paderborn, Germany}
\author{Artur V. Trifonov}
\affiliation{Spin Optics Laboratory, St. Petersburg State University, 198504 St. Petersburg, Russia}
\affiliation{Experimentelle Physik 2, Technische Universit\"at Dortmund, 44227 Dortmund, Germany}
\author{Polina R. Sharapova}
\affiliation{Department of Physics and Center for Optoelectronics and Photonics Paderborn (CeOPP), Paderborn University, D-33098 Paderborn, Germany}
\author{Jan Sperling}
\affiliation{Institute for Photonic Quantum Systems (PhoQS), Paderborn University, D-33098 Paderborn, Germany}
\affiliation{Department of Physics and Center for Optoelectronics and Photonics Paderborn (CeOPP), Paderborn University, D-33098 Paderborn, Germany}
\author{Dmitri R. Yakovlev}
\affiliation{Experimentelle Physik 2, Technische Universit\"at Dortmund, 44227 Dortmund, Germany}
\affiliation{Dortmund Center for Advanced Exploration of Dynamics Across Limits Using Spectroscopy (DAEDALUS), Technische Universit\"at Dortmund, 44227 Germany}
\author{Elena~V.~Kolobkova}
\affiliation{ITMO University, 199034 St. Petersburg, Russia}
\affiliation{St. Petersburg State Institute of Technology, 190013 St. Petersburg, Russia}
\author{Maria~S.~Kuznetsova}
\affiliation{Spin Optics Laboratory, St. Petersburg State University, 198504 St. Petersburg, Russia}
\author{Marc A{\ss}mann}
\affiliation{Experimentelle Physik 2, Technische Universit\"at Dortmund, 44227 Dortmund, Germany}
\affiliation{Dortmund Center for Advanced Exploration of Dynamics Across Limits Using Spectroscopy (DAEDALUS), Technische Universit\"at Dortmund, 44227 Germany}
\author{Manfred Bayer}
\affiliation{Experimentelle Physik 2, Technische Universit\"at Dortmund, 44227 Dortmund, Germany}
\affiliation{Research Center FEMS, Technische Universit\"{a}t Dortmund, 44227 Dortmund, Germany}
\author{Torsten Meier}
\email{torsten.meier@uni-paderborn.de}
\affiliation{Institute for Photonic Quantum Systems (PhoQS), Paderborn University, D-33098 Paderborn, Germany}
\affiliation{Department of Physics and Center for Optoelectronics and Photonics Paderborn (CeOPP), Paderborn University, D-33098 Paderborn, Germany}
\author{Ilya A. Akimov}
\email{ilja.akimov@tu-dortmund.de}
\affiliation{Experimentelle Physik 2, Technische Universit\"at Dortmund, 44227 Dortmund, Germany}
\affiliation{Dortmund Center for Advanced Exploration of Dynamics Across Limits Using Spectroscopy (DAEDALUS), Technische Universit\"at Dortmund, 44227 Germany}

\begin{abstract}
Photon echo (PE) spectroscopy is a powerful technique for probing decoherence mechanisms and charge carrier dynamics in semiconductor systems. Beyond traditional coherence measurements, characterizing the photon statistics of the echo signal is important for assessing its potential in quantum information applications and understanding the underlying quantum mechanical processes. Here, we study the photon statistics of PE signals generated by excitons in ensembles of lead halide perovskite CsPbI$_3$ nanocrystals at cryogenic temperature of 2 K using continuous-variable quantum state optical tomography based on homodyne detection. Pronounced Rabi oscillations of the PE amplitude allow us to evaluate the statistics for various pulse areas in the excitation sequence. The damping of the oscillations with increasing pulse area is attributed to spatial excitation inhomogeneity and excitation-induced dephasing. Despite the large ensemble of optically addressed excitons, the efficiency of generated PE signals is low which is attributed to the complex energy level structure of excitons and non-radiative recombination channels in CsPbI$_3$ nanocrystals. We analyze the statistical characteristics of PE via the second-order correlation function $g^{(2)}(0)$ and the characteristic function for different combinations of the areas of the excitation pulses. Our results show that $g^{(2)}(0) = 1$, and the characteristic function of the PE signal corresponds to classical behavior.  The formation of photon echoes as well as $g^{(2)}(0) = 1$ at the echo time is reproduced by a quantized free-space multimode model. Despite the relatively low efficiency, the photon echo exhibits a high degree of coherence and minimal classical noise, consistent with Poissonian statistics.
\end{abstract}

\maketitle

\section{Introduction}

Metal halide perovskite nanocrystals (NCs) have recently emerged as a promising class of materials for optoelectronic and quantum optical devices~\cite{Dey2021}. The elementary optical excitations in lead halide perovskites are excitons, which determine their exceptional properties, including high photoluminescence quantum yields, tunable emission wavelengths, and strong light–matter interactions~\cite{Dey2021,Tsarev2025,Zaffalon2025,Kopteva2024,Kopteva2025,Kirstein2023,Tamarat2023}. Recent studies have demonstrated their potential for single-photon generation and strong nonlinear optical responses, making them a robust platform for exploring fundamental light–matter interactions and for possible applications in quantum optical devices~\cite{Ferrando2018,Raino2016,Zhu2022,Wang2016,Xu2016}.

Photon echo is considered as a powerful technique for probing coherent dynamics of excitons in perovskite semiconductors~\cite{Becker2018,Liu2021,Trifonov2022,Grisard2023,Trifonov2025arXiv}. It relies on resonant excitation of excitons by a sequence of short optical pulses, where the rephasing process restores the coherent polarization induced by the first pulse, resulting in the emission of a delayed echo pulse~\cite{Mukamel-book,Cundiff2008,Poltavtsev2018,Groll2025}. The photon echo technique allows the measurement of coherence times and evaluation of the homogeneous linewidth of exciton in ensembles of NCs, despite strong inhomogeneous broadening. In the context of perovskite nanocrystals, photon echo experiments revealed rich quantum dynamics such as quantum beats originating from exciton fine structure splitting~\cite{Liu2021,Trifonov2025arXiv}. 
However, beyond measuring the photon echo amplitude, analyzing and characterizing the photon statistics of the echo signal is of particular importance for assessing its suitability in quantum optical applications. In particular, for applications such as quantum memories, it is essential that the storage and retrieval process does not introduce additional noise or alter the nonclassical photon statistics of the input field. Furthermore, the statistical properties of the photon echo can provide a deeper understanding of the underlying quantum mechanical processes governing coherence and dephasing in the studied materials.

In the context of quantum optics, photon echoes deserve special attention as a manifestation of coherent light–matter interaction and collective emission phenomena. While photon echoes have been extensively discussed in the framework of quantum memory protocols in rare-earth–doped crystals and atomic systems~\cite{Lvovsky2009,Tittel2010,Moiseev2025}, they also provide a unique platform for exploring quantum optical effects at the level of weak and even single-photon fields~\cite{Ohlsson2003}. It has been shown that the conventional two-pulse photon echo is unsuitable for quantum memory applications due to strong quantum noise originating from spontaneous emission~\cite{Ruggiero2009,Ma2021}. Nevertheless, the underlying processes give rise to a variety of intrinsically quantum-optical phenomena, e.g., rephased amplified spontaneous emission~\cite{Beavan2012}, which can exhibit nontrivial temporal correlations and is expected to strongly influence the photon statistics of the emitted field. However, despite these developments, quantum tomography of the emitted fields in photon-echo protocols has received little attention as most studies have focused on single-photon excitation and photon-counting detection schemes.

In semiconductor systems, photon echo has been demonstrated up to the level of single quantum emitters, including self-assembled quantum dots~\cite{Borri2001,Langbein2005,Kasprzak2011,Raymer2013,Wigger2023}. These measurements were focused on the acquisition of phase-resolved classical fields using heterodyne detection and two-dimensional Fourier spectroscopy. A key advantage of semiconductor quantum emitters is the large oscillator strength of excitons, leading to short radiative lifetimes below 1~ns and enabling experiments at very high repetition rates, reaching up to 1~GHz~\cite{Evers2021}. This makes semiconductor platforms particularly attractive for systematic investigations of photon-echo statistics using continuous-variable quantum state tomography~\cite{Lvovsky2009_CQST, Roumpos2013,Luders2018,Trifonov2025OE}. In this context, homodyne detection represents an excellent tool for evaluating the statistical and correlation properties of photon-echo fields, offering high temporal resolution together with fast operation enabled by high repetition rates.

In this work, we modify our previously developed real-time optical homodyne detection scheme~\cite{Trifonov2025OE} and employ it to investigate the photon statistics of photon echo signals generated in ensembles of quantum emitters based on perovskite CsPbI$_3$ nanocrystals. The paper is organized as follows. Section~\ref{sec:S&E} introduces the investigated sample (\ref{sec:Sample}) and two-pulse photon echo experimental setup (\ref{sec:Setup}). The details on homodyne detection and evaluation of phase-resolved quadratures are described in Sec.~\ref{sec:HomoD}. The main experimental results in Sec.~\ref{sec:R&D} are divided in two parts. First, in Sec.~\ref{sec:Rabi}, we demonstrate Rabi oscillations in the PE signal, a result that has not yet been reported for perovskite NCs. The damping of the Rabi oscillations is partly caused by spatial inhomogeneity of the excitation (transverse beam profile and the finite optical thickness of the sample) as well as by excitation-induced dephasing, as evidenced by the dependence of the optical coherence time $T_2$ on the excitation strength. Second, in Sec.~\ref{sec:Stat}, we evaluate the correlation and characteristic functions of the PE signal for excitation pulses of different areas by analyzing a set of phase-sensitive quadratures acquired via homodyne detection. Sec.~\ref{sec:Modeling} presents the theoretical modeling by a quantized free-space multimode description. The discussion in Sec.~\ref{sec:Disc} highlights the main results, namely the classical behavior of two-pulse photon echoes under excitation with laser pulses and negligible contribution of spontaneous emission due to use of homodyne detection with pulse duration much shorter than the exciton lifetime. Here, we also address the origin of low efficiency of the photon echo in CsPbI$_3$ NCs. A conclusion and outlook are given in Sec.~\ref{sec:Conc}.

\section{Sample and Experimental Setup}
\label{sec:S&E}

\subsection{Sample}
\label{sec:Sample}

\begin{figure}[t]
\includegraphics[width=\linewidth]{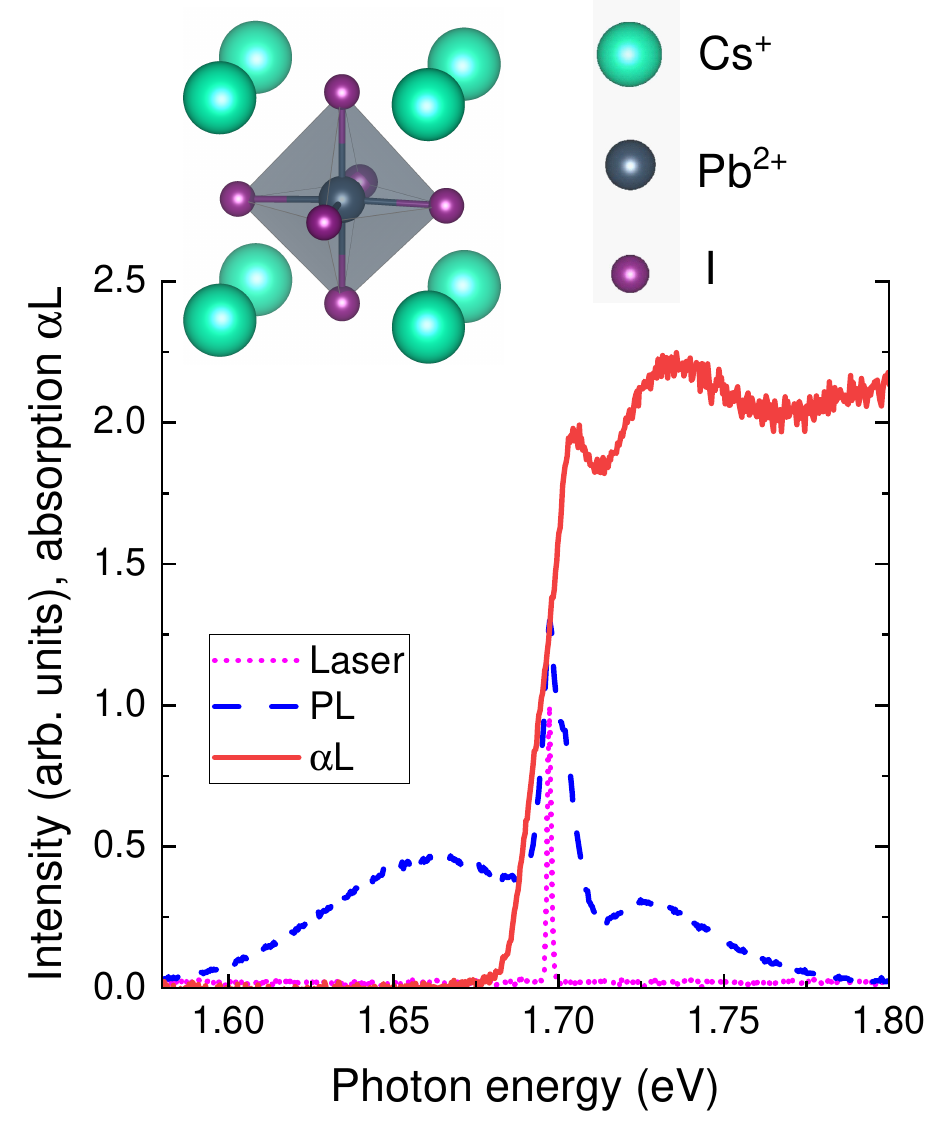}
\caption{Photoluminescence (PL, dashed blue line) and absorption coeficient $\alpha(h\nu) L$ (solid red line) spectra of CsPbI$_3$ nanocrystals measured at $T=2$~K. The PL spectrum was measured in backward emission geometry under excitation with photon energy of 2.33~eV. The laser spectrum used in the PE measurements is shown with the dotted magenta line. Inset shows the CsPbI$_3$ perovskite crystal structure.}
\label{fig:PL}
\end{figure}

\begin{figure*}[t]
\includegraphics[width=0.8\linewidth]{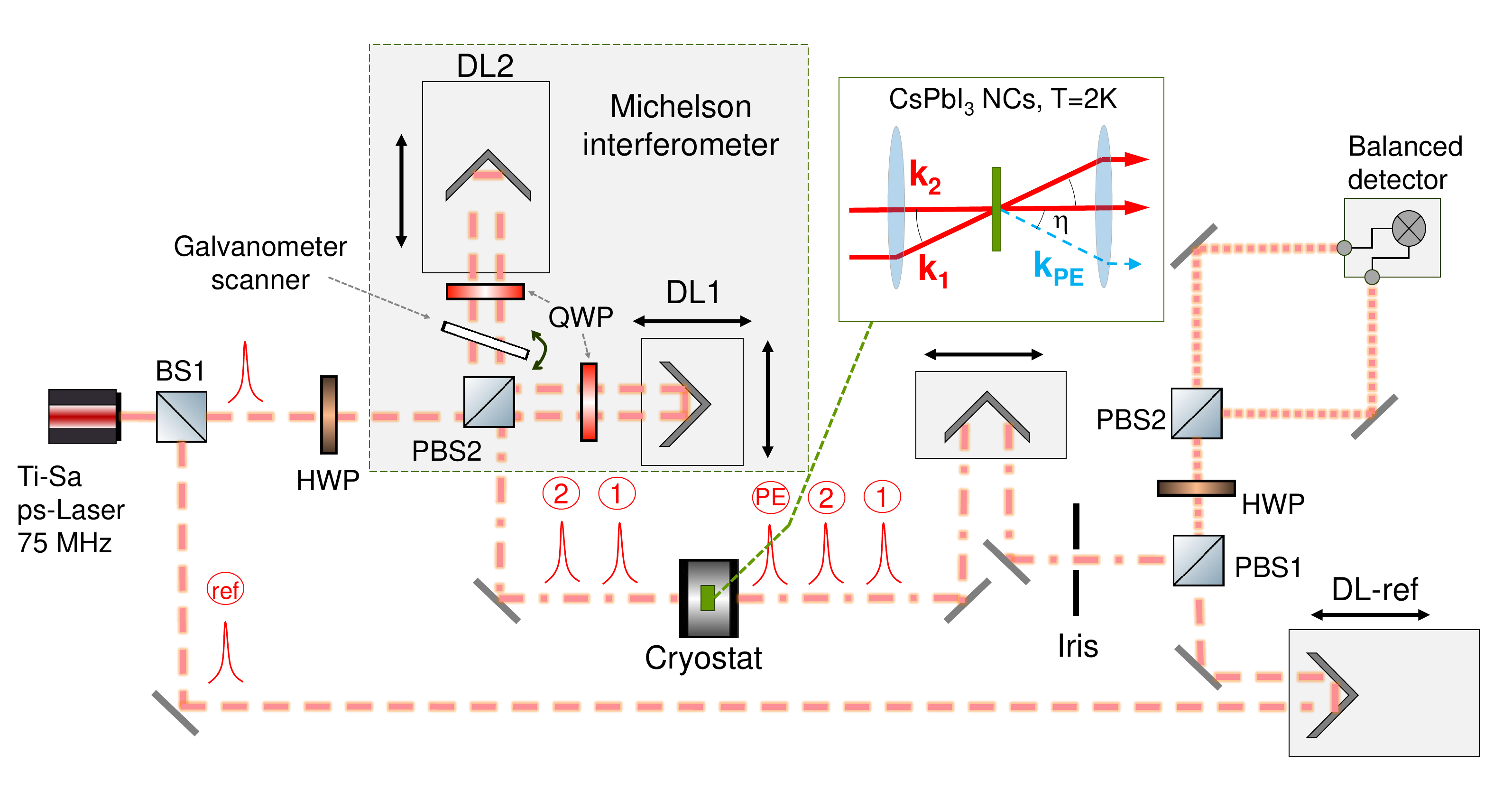}
\caption{Schematic presentation of the experimental setup for measuring and characterizing the photon echo. Notations: BS -- beam-splitter; PBS -- polarizing beamsplitter; DL -- delay line, HWP -- half waveplate, QWP -- quarter waveplate. PE -- photon echo, ref -- reference. The inset shows a close-up of the sample in the cryostat with the excitation and four-wave mixing beam geometry.}
\label{fig:SetUP}
\end{figure*}

The investigated sample contains CsPbI$_3$ nanocrystals embedded in fluorophosphate Ba(PO$_3$)$_2$-AlF$_3$ glass. The sample (reference number EK8) was synthesized by rapid cooling of a glass melt enriched with the components needed for the perovskite crystallization. More details on the preparation method of the sample are given in Refs.~\onlinecite{Kolobkova2021,Nestoklon2023}. The average diameter of the nanocrystals is about 10 to 15~nm. The sample thickness $L$ is about 200~$\mu$m. The absorption and photoluminescence (PL) spectra of the sample measured at the temperature of $T=2$~K are shown in Fig.~\ref{fig:PL}. The absorption band is characterized by an edge located at a photon energy of about $h\nu=1.70$~eV. The PL comprises a sharp peak approximately at the same energy (1.697 eV), which is superimposed on a broader peak in the spectral range of 1.6–1.8~eV. Such inhomogeneous spectral broadening arises from variations in the size of the nanocrystals. Moreover, the presence of narrow and broad PL bands is attributed to two sub-ensembles of NCs with different sizes. The narrow absorption peak originates, possibly from one of the sub-ensembles of NCs with a narrower size distribution. 
Subsequently, the laser used in PE experiments was tuned to the energy 1.697~eV (see Fig.~\ref{fig:PL}). 

\subsection{Photon Echo Setup}
\label{sec:Setup}

The experimental setup is illustrated in Fig.~\ref{fig:SetUP}. The setup includes a Ti:sapphire laser (Mira-900), which is tunable in the range of 700–990~nm and pumped by a 532~nm continuous-wave laser. The laser generates picosecond pulses with a duration of about 3~ps at a repetition frequency of 75~MHz and is tuned to a photon energy of 1.697 eV. We intentionally use spectrally narrow optical pulses with a full width at half maximum of about 0.6~meV in order to address signals related to the zero-phonon exciton line with long-lived coherence on the order of hundreds of picoseconds. This allows us to establish Rabi flopping of excitons in nanocrystals, which is otherwise difficult when using femtosecond optical pulses, where multiple exciton-polaron states have to be considered~\cite{Trifonov2025arXiv}. 

The laser beam is split by beam splitters into two excitation pulses using a Michelson interferometer (labeled as 1 and 2 in Fig.~\ref{fig:SetUP}) and one reference pulse (labeled as ref in Fig.~\ref{fig:SetUP}). The delay time between optical pulses is controlled using mechanical translation stages (DL$_1$, DL$_2$, DL$_{\rm ref}$). Specifically, the second pulse is delayed by an interval $\tau_{12}$ relative to the first pulse, and the reference pulse is delayed by $\tau_{\rm ref}$ relative to the first pulse.
We can switch between the co-linear and non-colinear geometry by transversely shifting the retroreflector on DL$
_1$, i.e. by changing the distance between beams exiting the interferometer. In this work, the non-colinear geometry was used to minimize stray light from the excitation pulses. The first pulse is transversely shifted by 3~mm relative to the second pulse. This shift results in a difference in their incidence angles of $\eta=0.03$~rad at the sample, after passing through the focusing lens placed before the cryostat with the sample (see the inset of Fig.~\ref{fig:SetUP}). Note that the second beam arrives under normal incidence, and, therefore, $\eta$ is defined by the angle of incidence of the first pulse. The PE signal was measured in the transmission geometry and collected in the phase-matching direction defined by $\mathbf{k}_{\rm PE}=2\mathbf{k}_2-\mathbf{k}_1$.  In this case, when perfect phase matching (co-linear geometry), corresponding to $\Delta k = 0$, is not satisfied, the resulting decrease in the efficiency of the PE generation can be estimated using 
$I/I_0=\mathrm{sinc}^2\left(\frac{\Delta k L}{2}\right)=0.62$, where $\Delta k L = 2\pi n L \eta^2/ \lambda =2.34$ is the phase mismatch ($\lambda$ is the excitation wavelength, and $n=1.5$ is the refractive index of the fluorophosphate glass)~\cite{Boyd-book}.

Measurements were performed in the polarization configuration, where the first and second pulses are linearly cross-polarized, while the detection polarization is parallel to the polarization of the first pulse. This scheme allows for further suppression of the stray light from the second pulse by filtering the signal using a polarizer in the detection path. In this configuration, however, the PE signal originating from excitons is typically weaker. For exciton transitions, signals are generally observed only in the co-polarized configurations, consistent with the expectation for a V-type energy scheme~\cite{Poltavtsev2019}. In the cross-polarized configuration, the exciton contribution becomes observable due to exciton fine-structure splitting, random orientation of nanocrystals in the ensemble, and dephasing arising from fluctuations of the fine-structure splitting among NCs of different sizes~\cite{Liu2021,Han2022}. As recently demonstrated for similar CsPbI$_3$ nanocrystals~\cite{Trifonov2025arXiv}, these effects allow one to observe the PE signal even in cross-polarized configuration, albeit with an electric field amplitude about five times smaller than in the co-polarized case.

The sample was placed in a helium bath cryostat and kept at a temperature of $T=2$~K. The excitation pulses are focused onto the sample within a spot diameter of approximately 50~$\mu$m using a convex spherical achromatic doublet lens with a focal length of 100~mm.  From these parameters, we can estimate the number of NCs addressed in the experiment. The NC density in the sample is approximately 10$^{15}$~cm$^{-3}$~\cite{Nestoklon2023}. The illuminated volume amounts to $V=4\times10^{-7}$cm$^3$. However, due to the strong inhomogeneous broadening, only a small fraction of NCs is resonant with the excitation pulse. This fraction can be estimated from the ratio of the laser spectral width $\approx0.6$~meV (full width at half maximum, FWHM) to the PL linewidth (FWHM $\approx 50$~meV) which is about $10^{-2}$. Thus, we estimate a number of resonantly addressed NCs in the order of $N_{\rm X} \approx 5 \times 10^6$.

The phase of the second pulse is modulated using a single-axis scanning galvanometer that rotates a glass plate placed in one arm of the Michelson interferometer at an incidence angle of approximately $20^\circ$. The glass plate undergoes angular modulation following a sawtooth waveform, with an amplitude of approximately  $0.5^\circ$ at a frequency of 100~Hz (see Fig.~\ref{fig:homoDet}(a)). Each half-period of the oscillation changes the phase by about $11\pi$, resulting in a linear phase sweep and a modulation of the cross-correlation between the second laser beam and the reference at approximately 1.1 kHz. Note that the phase change is nonlinear due to the acceleration and deceleration between the deflection points in the glass plate movement. Therefore, the acquisition window was set to span a phase interval of $2\pi$ within the central region of the glass-plate motion. In this defined region, the phase change can be considered linear, which simplifies the processing of the acquired data.

Phase modulation serves two key purposes in our experiment. First, modulating the phase of the second pulse also modulates the phase of the PE, and this enables us to perform a phase sensitive measurement of the PE quadrature, which will be described in detail below in Sec.~\ref{sec:HomoD}. Second, it allows us to distinguish between different contributions of optical fields in the detection channel. In particular, contributions from the coherent PE signal, coherent stray light and incoherent signals such as spontaneous emission can be evaluated separately using the phase modulation (phase cycling) approach. Coherent contributions can be distinguished according to the frequency phase-matching condition, the phase of the photon echo is given by the combination of the phases of the excitation pulses $\varphi_{\rm PE}=2\varphi_2-\varphi_1$. As a result, modulating the phase of the second pulse at frequency $f$ leads to different modulation frequencies of the various signals: the PE signal
and coherent stray light contributions are modulated at frequencies $f_{\rm PE}= 2f$ and $f$, respectively, while there is no modulation of the signal arising due to spontaneous emission. In some measurements, where only the amplitude of the PE signal is measured, we employ filtering at $2f$, and acquire the PE signal amplitude through a lock-in detection algorithm.

In the detection section of the setup, the PE signal is combined with a strong reference beam at a polarizing beam splitter (PBS1 in Fig.~\ref{fig:SetUP}), and the polarization of both signals is rotated by 45$^\circ$. The intensity of the reference beam is 1.5~mW, where, at this power level, the shot noise of the reference field dominates over the electronic noise from the detector and electronics. The mixed signal is then split equally into two channels before reaching a balanced photodetector. This configuration allows us to measure the cross-correlation between the reference pulse and the PE signal. 

\subsection{Homodyne Detection}
\label{sec:HomoD}

We use the homodyne detection scheme described in Ref.~\onlinecite{Luders2018}. A weak input signal is mixed with a strong local oscillator (in this work the reference is used as local oscillator). These signals are typically combined using a beam splitter, which induces a $\pi$ phase shift between the signals at the output. After the beam splitter, the combined signal is detected by photodetectors. The balanced photodetection provides the difference in the intensities $n_-$ measured at its different channels. The signal $n_-$ is directly proportional to the quadrature of the weak signal, $q_{\varphi}$, with the proportionality factor determined by the amplitude of local oscillator, $\alpha_{\rm LO}$:
\begin{equation}
{{\hat{q}}_\varphi} = \frac{{{{\hat{n}}_- }}}{{\sqrt 2 {\alpha }_{\textrm{LO}}}}, 
\end{equation}
where $\alpha_{\mathrm{LO}}$ is the amplitude of the local oscillator.

These quadratures can then be used to calculate various characteristics of the light field, including photon statistics and the second-order correlation function $g^{(2)}(0)$. The second-order correlation function is a measure of the statistical properties of a light field. For a coherent state, we have $g^{(2)}(0) =1$, while for non-classical light, such as a single-photon emitter, we find $g^{(2)}(0) <1$, and, in the case of a thermal state, we obtain $g^{(2)}(0)>1$. In our experiment, we calculate $g^{(2)}(0)$ using the following quadrature-based formula~\cite{Roumpos2013}:
\begin{equation}
g^{(2)}(\tau = 0, t)
=
\frac{\langle \hat{a}_s^\dagger \hat{a}_s^\dagger \hat{a}_s \hat{a}_s \rangle}
     {\langle \hat{a}_s^\dagger \hat{a}_s \rangle^2}
=
\frac{4\langle \hat{q}_\varphi^{\,4} \rangle
      - 12\langle \hat{q}_\varphi^{\,2} \rangle
      + 3}
     {6\left(\langle \hat{q}_\varphi^{\,2} \rangle - \tfrac{1}{2}\right)^2}.
\label{eq:g2}
\end{equation}
The averaging of	 quadratures occurs over a large array of accumulated experimental values~\footnote{Note that the definition of quadratures is to some degree arbitrary and changing their magnitude correpsonds to a redefinition of the value of noise of the vacuum state. Therefore the constant factors arising in this equation depend strongly on this choice and the exact measurement scheme used and therefore differ significantly in the literature.}. This expression is valid only for phase-averaged measurements, which allows for the assumption that $\langle q_\varphi \rangle =0$. The condition is naturally satisfied when the signal and the local oscillator originate from independent sources, resulting in a random relative phase. However, the photon echo signal retains a well-defined phase relationship to the local oscillator, as both are derived from the same laser source. To meet the phase-averaging requirement in our case, the following data processing method was implemented.

\begin{figure}[t]
\includegraphics[width=\linewidth]{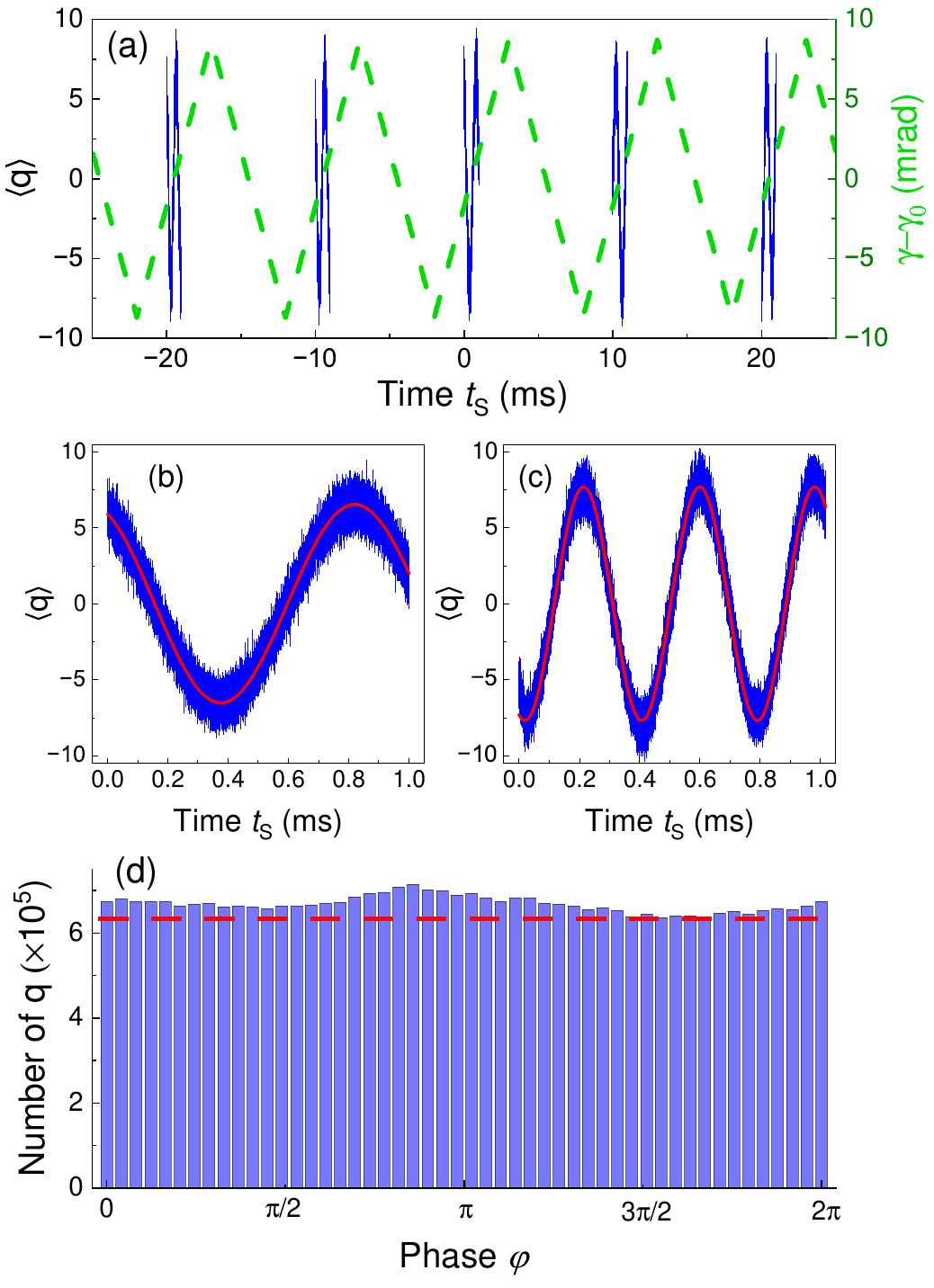}
\caption{(a) The blue curve represents a cross-correlation measurement between the reference and attenuated second excitation pulse (about 20 photons per pulse). The green curve is the corresponding deviation angle of the oscillating glass plate around the incidence angle $\gamma_0=0.27$~rad. Each segment (b) represents about $7\times10^4$ quadrature points collected at the center of the glass plate’s linear ramp. The blue curve shows the experimental data and the red curve is the sinusoidal fitting curve used to determine the phase for each quadrature point. (c) The segment shape for the PE signal, illustrating the double-frequency modulation compared to single frequency modulation for the excitation laser pulse shown in (b). (d) A histogram shows the initial unequal distribution of the number of quadratures, and the red dashed line illustrates the level of the distribution of quadrature samples after truncation. }
\label{fig:homoDet}
\end{figure}

Figure~\ref{fig:homoDet} illustrates the data acquisition and processing method, showing a cross-correlation measurement between the reference and a strongly attenuated second excitation pulse (about 20 photons per pulse) serving as the signal. The data acquisition is conducted in discrete segments, with each segment corresponding to about $7\times10^4$ quadrature points collected at the center of the glass plate’s linear ramp. Figure~\ref{fig:homoDet}(a) (blue curve) shows five segments of an acquired data sample and the corresponding deviation angle of the oscillating glass plate (green curve) in real time. For each segment, a sinusoidal function of the form $A\sin(2\pi f t_s + \delta)$ is fitted to the data to extract the phase information, where $t_s$ is the time stamp (see red curve  in Fig.~\ref{fig:homoDet}(b)). In the case of PE, the fitting function has the form $A\sin(4\pi f t_s+\delta)$ (see Fig.~\ref{fig:homoDet}(c)). Each measured quadrature value is then assigned to a corresponding phase $\varphi$ between 0 and $2\pi$. Due to the phase fluctuations of the reference, different data segments cover different phase ranges. Therefore, as shown in Fig.~\ref{fig:homoDet}(d), different phase regions accumulate different numbers of quadratures. To achieve uniform phase sampling, we randomly truncate the number of quadrature points in the overrepresented regions, ensuring a homogeneous distribution of quadrature measurements across all phases (Fig.~\ref{fig:homoDet}(d) red dashed red line).

\begin{figure}[t]
\includegraphics[width=\linewidth]{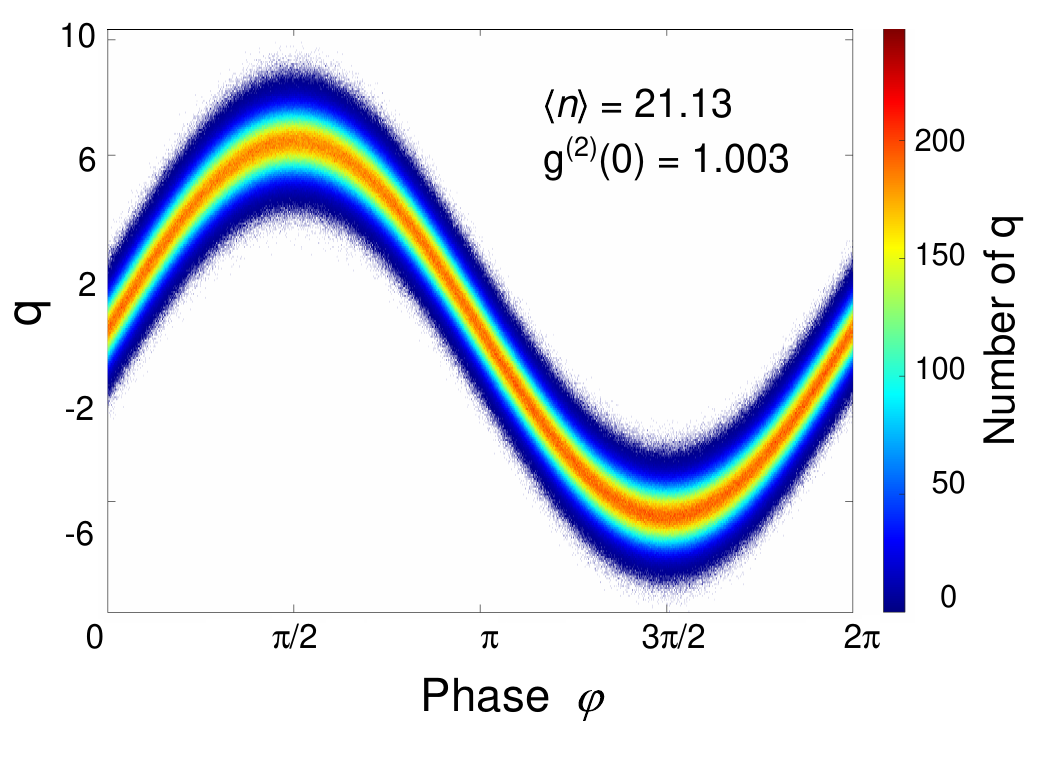}
\caption{The histogram of phase-resolved quadratures for a highly attenuated second excitation laser pulse. The figure is based on $3\times 10^7$ acquired data points. }
\label{fig:quadratures}
\end{figure}

By using an equal number of quadratures across all phases, the condition for phase-averaged measurements $\langle q_\varphi \rangle =0$ is effectively satisfied, allowing us to apply Eq.~\eqref{eq:g2}. After this step, the distribution of quadrature values as a function of phase can be plotted. Figure~\ref{fig:quadratures} shows an example of acquired quadratures for the attenuated second excitation laser pulse. By applying Eq.~\eqref{eq:g2} on the quadrature set in Fig.~\ref{fig:quadratures}, we obtain a value of $g^{(2)}(0) =1.003 \pm 0.0001$ for the laser pulse which agrees with Poisson statistics expected for classical laser source. For phase randomized quadratures, we also calculate the average photon number per pulse $\langle n \rangle = \langle q^2 \rangle -0.5=21.13$, where we used the  expectation value of all squared quadratures. This value is in agreement with $\alpha_{S}^2/2$ obtained from sinusoidal fit with amplitude $\alpha_S=6.50$ in Fig.~\ref{fig:quadratures}. 

\section{Results and Discussion}
\label{sec:R&D}

\subsection{Rabi Oscillations of the Photon Echo}
\label{sec:Rabi}

\begin{figure}[t]
\includegraphics[width=0.9\linewidth]{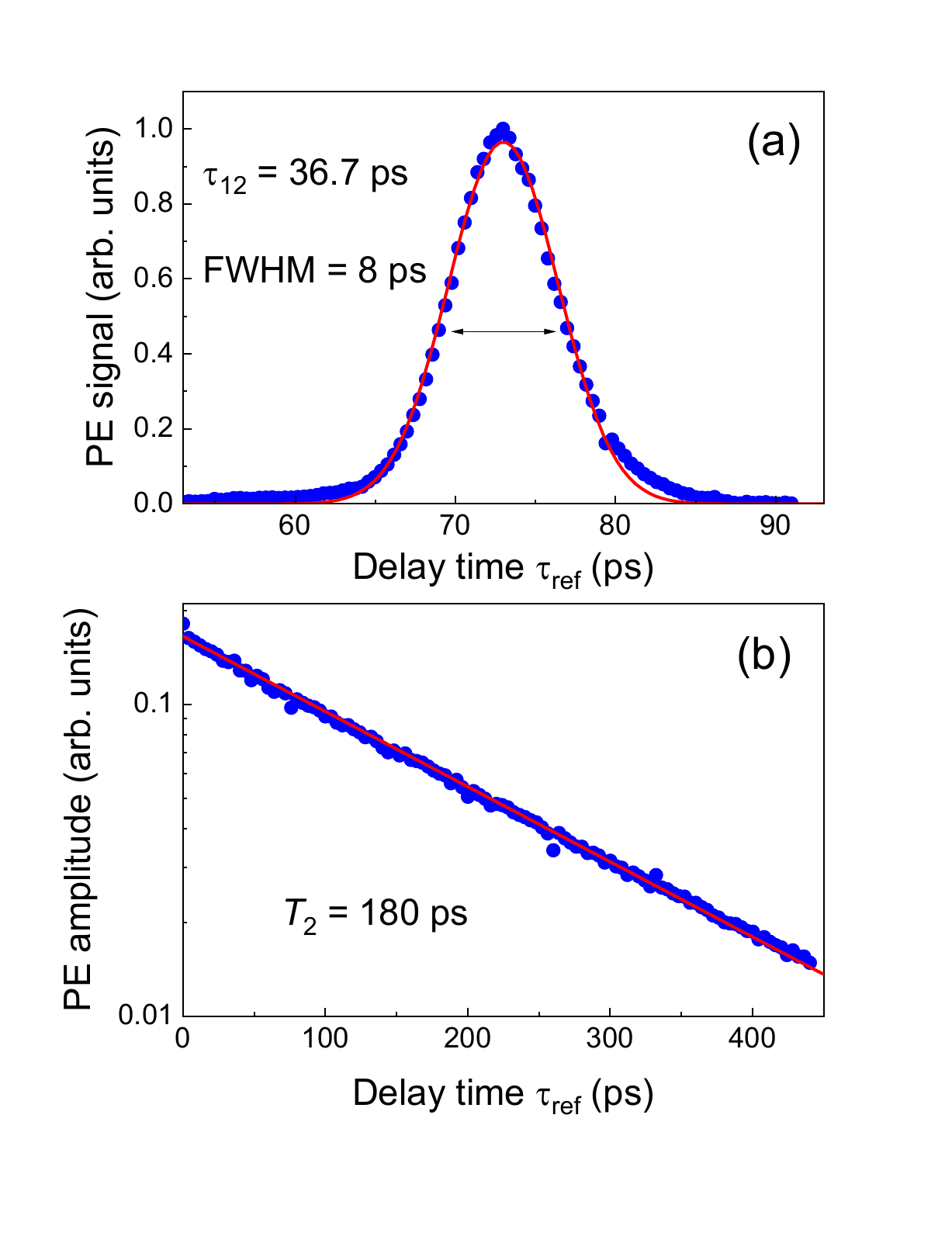}
\caption{The PE temporal profile (a) and PE amplitude decay (b) measured for $h\nu= 1.697$~eV, $T=2$~K, $\tau_{12}= 36.7$~ps, $A_1=4.5$~pJ$^{1/2}$, $A_2=9.0 $~pJ$^1/2$. Dots correspond to experimental data and lines to the fit with Gaussian (a) and exponential (b) functions.}
\label{fig:PE-profile}
\end{figure}

Scanning the time delay between the first pulse and the reference pulse $\tau_{\rm ref}$, while keeping delay time $\tau_{12}$ constant, constitutes a measurement of the temporal profile of the PE pulse. Figure~\ref{fig:PE-profile}(a) shows the temporal profile of the PE pulse at a delay time of $\tau_{12} = 36.7$~ps. The symbols represent experimental data, while the red curve is a Gaussian fit with a full width at half maximum (FWHM) of $\sigma_{\rm PE} \approx 8$~ps. The measured signal represents the cross-correlation between the optical fields of the PE and reference pulses, i.e. the convolution of these pulses. The width of the PE profile therefore reflects the spectral broadening of the excited ensemble, which is determined by the spectral width of the laser. Assuming Gaussian pulse shapes, we estimate the FWHM duration of the PE intensity pulse as $\sqrt{\sigma_{\rm PE}^2/2 - \tau_d^2} = 4.8$~ps, where $\tau_d = 3$~ps is the laser pulse duration. This is consistent with the regime where the inhomogeneous broadening of the ensemble is substantially broader than the laser pulse spectrum. 

A simultaneous scan of $\tau_{12}$ and $\tau_{\rm ref}=2\tau_{12}$ allows one to measure the decay of the PE amplitude and evaluate the exciton coherence time $T_2$. These data are shown in Fig.~\ref{fig:PE-profile}(b). Fitting the experimental curve in Fig.~\ref{fig:PE-profile}(b) with an exponential function $\propto \exp{(‒2\tau_{12}/T_2)}$ (red curve) yields $T_2 = 180$~ps. The value is in agreement with the coherence times measured in the range from 150 to 330~ps in similar CsPbI$_3$ NCs~\cite{Trifonov2025arXiv}. For higher excitation levels, the decay of the PE amplitude becomes faster and non-monotonic indicating the importance of excitation induced dephasing, discussed  at the end of this section (see also Appendix~\ref{sec:A:EID}).  

\begin{figure}
\includegraphics[width=0.75\linewidth]{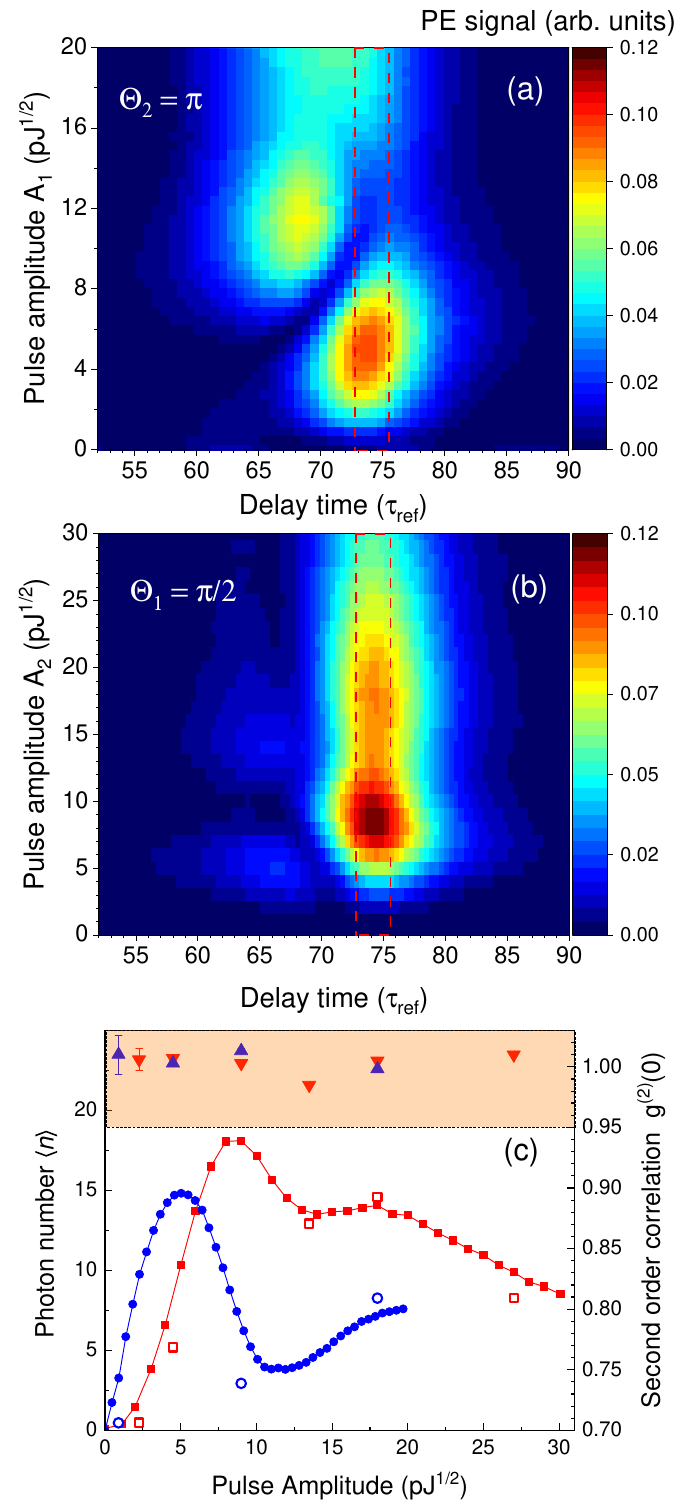}
\caption{(a) Rabi oscillations of PE amplitude. Measurements of PE transients (a) as a function of pulse amplitude $A_1$ at $A_2 = 9$~pJ$^{1/2}$ corresponding to pulse area $\Theta_2\approx \pi$ and (b) as a function of pulse amplitude $A_2$ at $A_1 = 4.5$~pJ$^{1/2}$ corresponding to pulse area $\Theta_1\approx \pi/2$.  (c) PE signal at $\tau_{\rm ref} = 2\tau_{12} = 73.4$~ps as a function of pulse anmplitudes $A_1$ and $A_2$. The data are evaluated within the time window of 3~ps centered at $2\tau_{12}$, as indicated by red rectangles in panels (a) and (b). Circles and squares correspond to variation of $A_1$ and $A_2$, respectively. Open symbols correspond to the number of photons evaluated in the homodyne measurement ($\tau_{\rm ref} = 2\tau_{12} = 73.4$~ps). Triangles correspond to the right axis and show the results of second-order correlation measurements summarized in Table~\ref{tab:g2}.}
\label{fig:Rabi}
\end{figure}

In order to measure Rabi oscillations, PE transients similar to the one shown in Fig.~\ref{fig:PE-profile}(a) are recorded as a function of the pulse area of one of the two excitation pulses, $\Theta_i$ ($i = 1,2$), while the area of the other pulse is kept constant. The pulse area is defined as $\Theta = \int_{-\infty}^{+\infty}\Omega_R(t)dt$, where $\Omega_R(t)=d_{12}E(t)/\hbar$ is the Rabi frequency with $E(t)$ being the time-dependent electric field amplitude of an 
optical pulse, and $d_{12}$ is the dipole matrix element of the optical transition. The area of the pulses is proportional to the pulse amplitudes $A_i$, which we define as the square root of the pulse energy. The variation of amplitudes $A_i$ is accomplished by attenuation of their intensities using half-wave and quarter-wave plates in combination with PBS as part of the Michelson interferometer in Fig.~\ref{fig:SetUP}. By scanning the areas of the incident pulses and measuring the temporal profiles of the photon echo, we observe Rabi oscillations displayed as two-dimensional images in Fig.~\ref{fig:Rabi}.

When the first pulse amplitude $A_1$ is varied at $A_2 = 9.0$~pJ$^{1/2}$, the PE transients reveal a complex two-dimensional picture with a split PE profile similar to that observed recently in a CdTe/(Cd,Mg)Te single quantum well and (In,Ga)As quantum dots~\cite{Poltavtsev2017,Poltavtsev2016}. The first maximum of the Rabi oscillations for $A_1 = 4.5$~pJ$^{1/2}$ corresponding to $\Theta_1 \approx \pi/2$ is centered at $2\tau_{12}\approx 73.4$~ps and is well described by a single pulse of Gaussian shape. However, the second maximum at $A_1 = 11.5$~pJ$^{1/2}$ is advanced by 5.2 ps. The third maximum of Rabi oscillations at $A_1 > 16$~pJ$^{1/2}$ appears to be also shifted by 3 ps (see Fig.~\ref{fig:Rabi}(a)). This behavior is well established in the case of photon echoes with strong inhomogeneous broadening of optical transitions that is larger than the spectral width of the laser. In this case, there is a considerable dephasing during the action of the laser pulse which depends on its area. As a result, the temporal profile of echo experiences significant variations in time as studied in detail in Refs.~\onlinecite{Poltavtsev2016, Kosarev2020, Grisard2023-Sorting}.  

When the amplitude of the second pulse $A_2$ is varied at $A_1 = 4.5$~pJ$^{1/2}$, the PE maximum has no detectable shift, as can be seen in Fig.~\ref{fig:Rabi}(b), in agreement with previous studies in InGaAs quantum dots~\cite{Kosarev2020,Grisard2023-Sorting,Grisard2022}. Here, however, only the first maximum of Rabi oscillations at $A_2 = 9.0$~pJ$^{1/2}$ ($\Theta_2 \approx \pi$) is pronounced, while the second one around $A_2 = 18$~pJ$^{1/2}$ ($\Theta_2 \approx 2\pi$) is weak. This is more easily visible in Fig.~\ref{fig:Rabi}(c), where the cross sections of the two-dimensional plots of Rabi oscillations as a function of both pulse amplitudes are plotted.

We observe that an increase in the area of either excitation pulse leads to a damping of the photon echo signal. One of the reasons for such damping is the spatial inhomogeneity of the excitation, which includes both the Gaussian beam profile in the plane of the sample~\cite{Poltavtsev2017} and notable absorption along the propagation direction due to the finite optical thickness of the sample (see Fig.~\ref{fig:PL})~\cite{Ruggiero2009}. In addition, excitation-induced dephasing plays an important role in our data, as manifested by the power dependence of the photon echo decay curves, where $T_2$ decreases non-monotonically from 210 to 100~ps, as discussed in Appendix~\ref{sec:A:EID}. 

Using the Rabi-oscillation calibration that $4.5~\mathrm{pJ}^{1/2}$ corresponds to a $\pi/2$ pulse area, together with the pulse duration, spot size, refractive index, and sample absorption, we estimate a transition dipole matrix element of $\mu \approx 15~\mathrm{D}$. This value is consistent in order of magnitude with transition dipole moments inferred from optical Stark effect measurements in related lead-halide perovskite nanocrystals~\cite{Lin2023NatNano,Li2020JPCL}.

\subsection{Second-Order Correlation and Characteristic Function}
\label{sec:Stat}

Here, we evaluate the results of quadrature measurements of the detected signals for different delay times $\tau_{12}$, $\tau_{\rm ref}$ , and pulse areas $A_1$, $A_2$. The quadratures are obtained using the procedure described in Section~\ref{sec:HomoD}. First, we calculate the correlation function $g^{(2)}(0)$ and photon number at the delay time corresponding to the PE maximum ($\tau_{\rm ref} = 2\tau_{12} = 73.4$~ps), which are summarized in Table~\ref{tab:g2}. The average number of detected photons per PE pulse is also shown in Fig.~\ref{fig:Rabi}(c) with open symbols. We notice in Table~\ref{tab:g2} that while the photon number changes in accordance with previously described Rabi flops when changing $\Theta_1$ or $\Theta_2$, the $g^{(2)}(0)$ is always $\approx 1$. This indicates that the photon echo signal from the perovskite nanocrystals sample exhibits classical photon statistics, which is consistent with the fact that the PE is coherent emission from the ensemble of emitters under excitation with a sequence of laser pulses. In addition, we measured the quadratures at the expected PE maximum when the first pulse is blocked. This results in $\langle q \rangle =0$ and $\langle q^2 \rangle = 0.5$, which corresponds to vacuum fluctuations. This value was obtained for all $\Theta_2$ presented in Table~\ref{tab:g2}. The results also coincide with the data measured at $\tau_{\rm ref} < 2 \tau_{12}$, i.e., away from the PE pulse for all the different combinations of $\Theta_1$ and $\Theta_2$ presented in Table~\ref{tab:g2}. 

\begin{table}
\caption{$g^{(2)}(0)$ and average photon number $\langle n \rangle$
measured for different pulse areas $\Theta_1$ and $\Theta_2$. $\tau_{\rm ref} = 2\tau_{12}=73.4$~ps.}
\label{tab:g2}
\begin{ruledtabular}
\begin{tabular}{cccc}
$\Theta_1$ & $\Theta_2$ & $g^{(2)}(0)$ & $\langle n \rangle$ \\ \hline
$\dfrac{\pi}{10}$ & $\pi$ & 1.010  & 0.47 \\ \hline
\multirow{6}{*}{$\dfrac{\pi}{2}$}
 & $\dfrac{\pi}{4}$   & 1.006 & 0.48 \\
 & $\dfrac{\pi}{2}$   & 1.007 & 5.17 \\
 & $\pi$              & 1.003 & 30.3 \\
 & $\dfrac{3\pi}{2}$  & 0.985 & 12.9 \\
 & $2\pi$             & 1.005 & 14.5 \\
 & $3\pi$             & 1.010  & 8.29 \\ \hline
$\pi$   & $\pi$ & 1.013  & 2.93  \\ \hline
$2\pi$  & $\pi$ & 0.998 & 8.26 \\
\end{tabular}
\end{ruledtabular}
\end{table}

Next, we calculate the characteristic function of the PE signal. The characteristic function is proved to be an effective criterion for detecting the deviation from classicality, particularly in cases where other conventional criteria fail~\cite{Kiesel2009,Ryl2017}. It is defined as the Fourier transform of the Glauber-Sudarshan P-function, which provides a quasi-probability distribution in phase space. The characteristic function $\Phi(\beta)$ can be estimated from quadrature measurements using \cite{Kiesel2009}
\begin{equation}
\overline{\Phi}(\beta) = \frac{1}{N} \mathrm{e}^{\frac{|\beta|^{2}}{4}} \sum_{j=1}^{N} \mathrm{e}^{\,i\,|\beta| q_j},
\label{eq:CharF}
\end{equation}
where $q_j$ are the quadratures measured at some specific phase $\varphi$, $N$ is the total number of quadratures for given $\varphi$.
The variance of the expectation value of the characteristic function is defined as
\begin{equation}
\sigma^{2}\!\left(\overline{\Phi}(\beta)\right) = \frac{1}{N} \left[ \mathrm{e}^{\frac{|\beta|^{2}}{2}} - \left|\overline{\Phi}(\beta)\right|^{2} \right].
\label{eq:CharFSigma}
\end{equation}
The nonclassical boundary is usually defined as the range of one standard deviation added to the characteristic function of the vacuum,
$\left|\Phi_{\mathrm{vac}}(\beta)\right| = 1$~\cite{Vogel-2000, Richter-2002}:
\begin{equation}
\left|\overline{\Phi}(\beta)\right| > 1 + S\,\sigma\!\left(\overline{\Phi}(\beta)\right).
\label{eq:boundary}
\end{equation}
This means that for a nonclassical state, this inequality should be satisfied at least at one point $\beta$ and with a minimum significance of $S = 1$.
Conversely, $|\Phi(\beta)|\leq 1$ holds true within the error margins for classical light. Note that higher-order criteria in terms of the characteristic function form a hierarchy of necessary and sufficient criteria for observing nonclassicality ~\cite{Vogel-2000, Richter-2002}.

\begin{figure}[t]
\includegraphics[width=\linewidth]{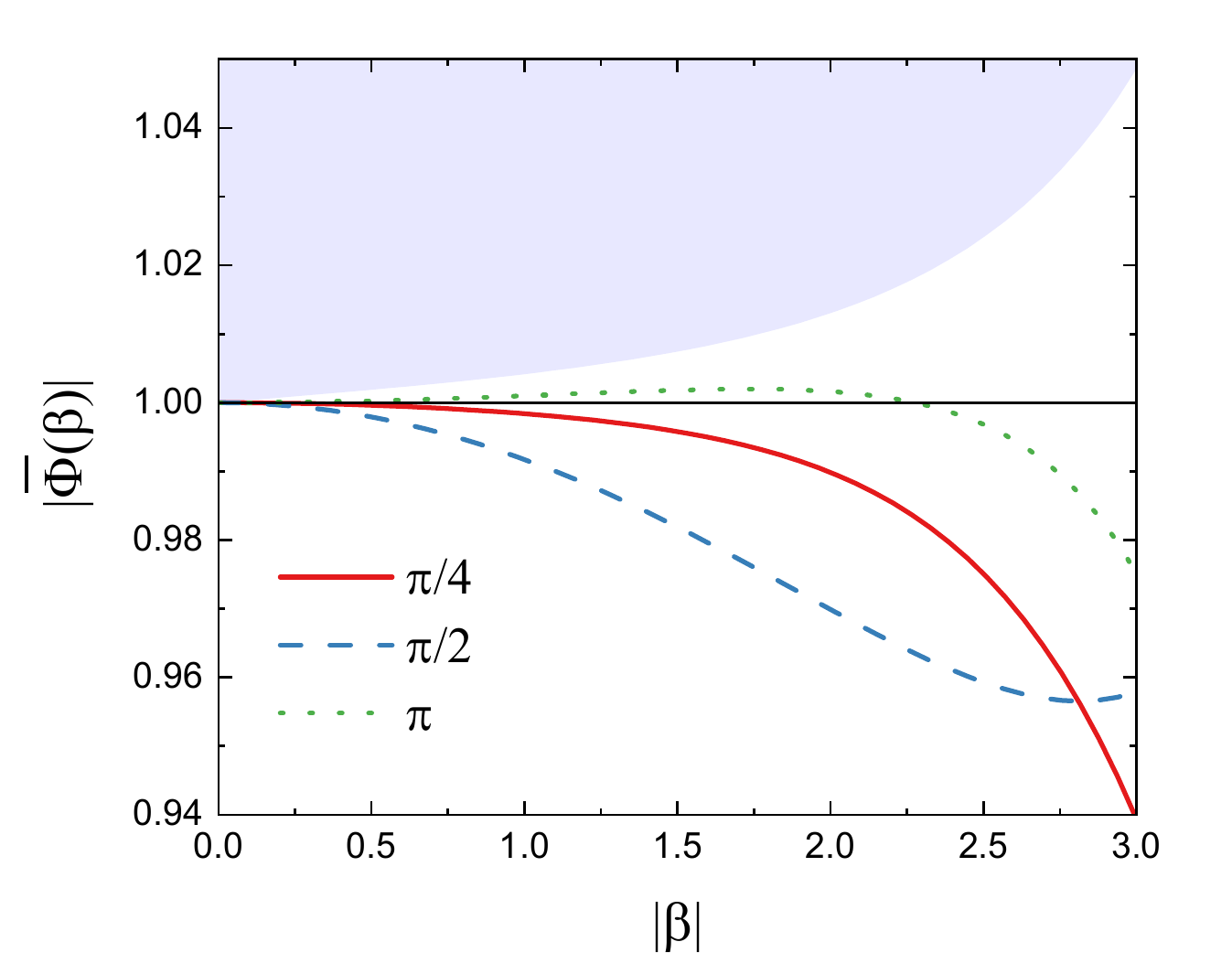}
\caption{Characteristic function $|\overline\Phi(\beta)|$ for photon echo signal obtained using Eq.~\eqref{eq:CharF}. Pulse areas correspond to $\Theta_1 = \pi/10$ and $\Theta_2= \pi$. The results are shown for different phases $\varphi = \pi/4$, $\pi/2$, and $\pi$. The shaded area corresponds non-classical region where a one standard deviation is added to the boundary $|\Phi(\beta)|=1$ according to Eq.~\eqref{eq:boundary}.}
\label{fig:CharF}
\end{figure}

In our analysis, Eq.~\eqref{eq:CharF} is applied to quadratures within a phase window of 0.04$\pi$ centered at $\varphi = \pi/4$, $\pi/2$, and $\pi$. This subset included approximately $3\times 10^4$ quadrature points. The characteristic function of the photon echo is found to be within the classical region as shown in Fig.~\ref{fig:CharF}. Similar results were obtained for all $\Theta_1$ and $\Theta_2$ indicated in Table~\ref{tab:g2}. This further supports the observations from the $g^{(2)}(0)$ measurements. The data indicate that, under the present experimental conditions, the photon echo exhibits purely classical features, while still preserving the coherence of the optical excitation.

\subsection{Theoretical Modeling}
\label{sec:Modeling}

We model the sample as a thin sheet of inhomogeneously broadened two-level systems (TLS) located at $z=0$, coupled to a multimode quantum light state. The matter is described by the coherence operator $\hat{\sigma}_-(\Delta,t)$ for emitters with detuning $\Delta$, and the field by the photon annihilation operator $\hat{a}(\omega,t)$, where $\omega$ denotes the difference to the central optical frequency in the rotating-wave approximation. The inhomogeneous distribution of the TLS is denoted by $f(\Delta)$ and the light-matter coupling by $g$. In the weak-excitation regime the inversion of the TLS is replaced by a scalar $s$ and therefore the Heisenberg equations for the operators are
\begin{align}
    \frac{\partial}{\partial t}{\hat{\sigma}}_-(\Delta,t)
    &=
    -i\Delta\hat{\sigma}_-(\Delta,t)
    +is\int d\omega\,g \hat a(\omega,t) ,
    \label{eq:model_sigma}
    \\
    \frac{\partial}{\partial t}{\hat a}(\omega,t)
    &=
    -i\omega\hat a(\omega,t)
    -i\int d\Delta\,f(\Delta) g^*
     \hat \sigma_-(\Delta,t) .
    \label{eq:model_a}
\end{align}
It should be noted that, unlike previous approaches  \cite{Ledingham2010, Moiseev2010}, we consider both an inhomogeneous  distribution of TLS and a multimode  free-space field. Therefore, the right-hand sides of  Eqs. \eqref{eq:model_sigma} and \eqref{eq:model_a} contain integrals over all field frequencies and detunings, respectively. This leads to a system of coupled integro-differential equations, which in the general case can only be solved numerically. 

The inhomogeneous distribution of the TLS is modeled by a Gaussian distribution
\begin{equation}
    f(\Delta) =
    \frac{N_{\mathrm{TLS}}}{\sqrt{2\pi}\sigma_\Delta}
    \exp\!\left[-\frac{\Delta^2}{2\sigma_\Delta^2}\right],
\end{equation}
with $\sigma_\Delta = \frac{\mathrm{FWHM}_\Delta}{2\sqrt{2\ln2}}$ and $N_{\rm TLS}$ is the number of TLS participating in the interaction.
The sample is initially unexcited, i.e., $\langle\hat\sigma_-(\Delta,t_0)\rangle = 0$.
The optical input is taken to be a multimode coherent state $|\alpha\rangle$,
defined by
\begin{equation}
    \hat{a}(\omega,t_0)\,|\alpha\rangle = \alpha(\omega)\,|\alpha\rangle,\label{eq:eigenvalue_coherent}
\end{equation}
with a Gaussian coherent spectral amplitude
\begin{equation}
    \alpha(\omega)
    =
    \alpha_0
    \frac{\exp[-\omega^2/(2\sigma_\omega^2)]}
    {\sqrt{\sigma_\omega}\,\pi^{1/4}} \exp[i\omega(t_1-t_0)].
    \label{eq:model_alpha_transfer}
\end{equation}
Here \(\sigma_\omega=\mathrm{FWHM}_\omega/(2\sqrt{\ln2})\), so that
\(\mathrm{FWHM}_\omega\) denotes the full width at half maximum of
\(|\alpha(\omega)|^2\). The phase factor $\exp[i\omega (t_1 - t_0)]$ sets the temporal offset of the incident pulse such that it impinges on the sample at $t=t_1$, while the simulation starts at $t=t_0$. In the following we use the experimental time convention \(t_1=0\) and the rephasing pulse is applied at \(t=\tau_{12}\), such that the photon echo appears near $t_{\rm echo}=2\tau_{12}$.

The numerical simulation is performed in two stages. For \(t_0\le t<\tau_{12}\) the system evolves with \(s=-1\). At \(t=\tau_{12}\) the rephasing pulse  is applied which is modeled as an ideal $\pi$-pulse. Thus, it conjugates the coherence operator $\hat \sigma_-(\Delta,\tau_{12})$, changes the inversion to $s=+1$, but leaves the optical field unchanged. The following evolution leads to the rephasing of the microscopic polarizations and the subsequent emission of the photon echo at $t_{\rm echo}=2\tau_{12}$.

Since Eqs.~\eqref{eq:model_sigma} and \eqref{eq:model_a} are linear, each stage can be written in terms of transfer functions \cite{PhysRevResearch.2.013371,Scharwald2023}, for which modeling details are given in Appendix~\ref{app:transfer_functions}. Moreover, the four-wave-mixing field is treated as a separate output mode. Since this mode contains the emitted photon echo in the present model, we refer to it as the echo mode in the following. After combining the two stages and the action of the $\pi$-pulse, the photon annihilation operator for the echo mode can be written as
\begin{align}
    \hat b(\omega,t)
    &=
    \int d\omega'\,
    B(\omega,\omega';t)\hat a^\dagger(\omega',t_0)
    \nonumber\\
    &\quad+
    \int d\Delta'\,
    Y(\omega,\Delta';t)\hat{\sigma}_-^\dagger(\Delta',t_0)
    +
    \hat b^{\mathrm{vac}}(\omega,t).
    \label{eq:model_b_operator}
\end{align}
The transfer functions \(B(\omega,\omega';t)\) and \(Y(\omega,\Delta';t)\)
contain the two-stage light-matter pathways contributing to the echo mode. $\hat b^{\rm vac}$ represents the contribution of the component that is the vacuum before the second-stage light-matter interaction,  and is defined in terms of transfer functions, see App.~\ref{app:transfer_functions}.

The considered positive-frequency field operator reads
\begin{equation}
\hat E^{(+)}(z,t)
=
\begin{cases}
\displaystyle
\int d\omega\,\mathcal{E}(\omega)
e^{ik(\omega)z}\hat a(\omega,t),
& t < \tau_{12}, \\[1.2em]
\displaystyle
\int d\omega\,\mathcal{E}(\omega)
e^{ik(\omega)z}\hat b(\omega,t),
& t \ge \tau_{12}.
\end{cases}
\label{eq:model_Eplus}
\end{equation}
where $\mathcal{E}(\omega)$ is a spectral prefactor, which in the present calculations is taken in arbitrary units. Equation~\eqref{eq:model_Eplus} is an expansion into delocalized plane-wave modes ($\propto~\exp[i k(\omega) z]$) with the linear dispersion relation $k(\omega) = \omega/c$. For \(t\ge\tau_{12}\), \(\hat E^{(+)}\) denotes the field in the echo mode.

To study the dynamics of the system, we introduce the normally-ordered intensity $I(z,t)$ and the macroscopic polarization $P(t)$:
\begin{align}
    I(z,t) &= \left\langle \hat E^{(-)}(z,t) \hat E^{(+)}(z,t)\right\rangle, \label{eq:model_intensity}\\
    P(t) &= \int d\Delta\, f(\Delta)\,\langle \hat{\sigma}_-(\Delta,t)\rangle,\label{eq:model_polarization}
\end{align}
with $\hat E^{(-)}(z,t) = \left[ \hat E^{(+)}(z,t) \right]^\dagger$.
The second-order correlation function for zero delay at the sample position $z=0$ can be simplified to
\begin{equation}
g^{(2)}(0,t)
=
\begin{cases}
1,
& t < \tau_{12}, \\[0.8em]
\displaystyle
2-
\left[
\frac{n_{\mathrm{coh}}(t)}
{n_{\mathrm{coh}}(t)+n_{\mathrm{inc}}(t)}
\right]^2,
& t \ge \tau_{12},
\end{cases}
\label{eq:model_g2_main}
\end{equation}
where $n_{\rm coh}(t)$ and $n_{\rm inc}(t)$ are coherent and incoherent contributions, respectively, whose definitions are given in App.~\ref{app:transfer_functions}. 
The second line of equation~\eqref{eq:model_g2_main} has the same form as the expression for displaced thermal light \cite{Lueders2021,Brune2025}, although it is obtained here for a multimode field.
For \(t<\tau_{12}\), the field mode is determined by the coherent input field and therefore \(g^{(2)}(0,t)=1\). For $t \ge \tau_{12}$, $g^{(2)}(0,t)$ 
is close to \(2\) at time for which the incoherent contributions dominate ($n_{\rm coh} \ll n_{\rm inc}$), whereas it approaches $1$ when the coherent echo dominates ($n_{\rm coh} \gg n_{\rm inc}$)
near $t=2\tau_{12}$.

The coupled equations for the transfer-functions, Eqs.~\eqref{eq:tf_mm} to \eqref{eq:tf_fm}, are solved numerically with the parameters listed in Table~\ref{tab:model_parameters}. Then, Eqs.~\eqref{eq:model_intensity}, \eqref{eq:model_polarization}, and \eqref{eq:model_g2_main} are evaluated.

\begin{figure}
    \centering
    \includegraphics[width=\columnwidth]{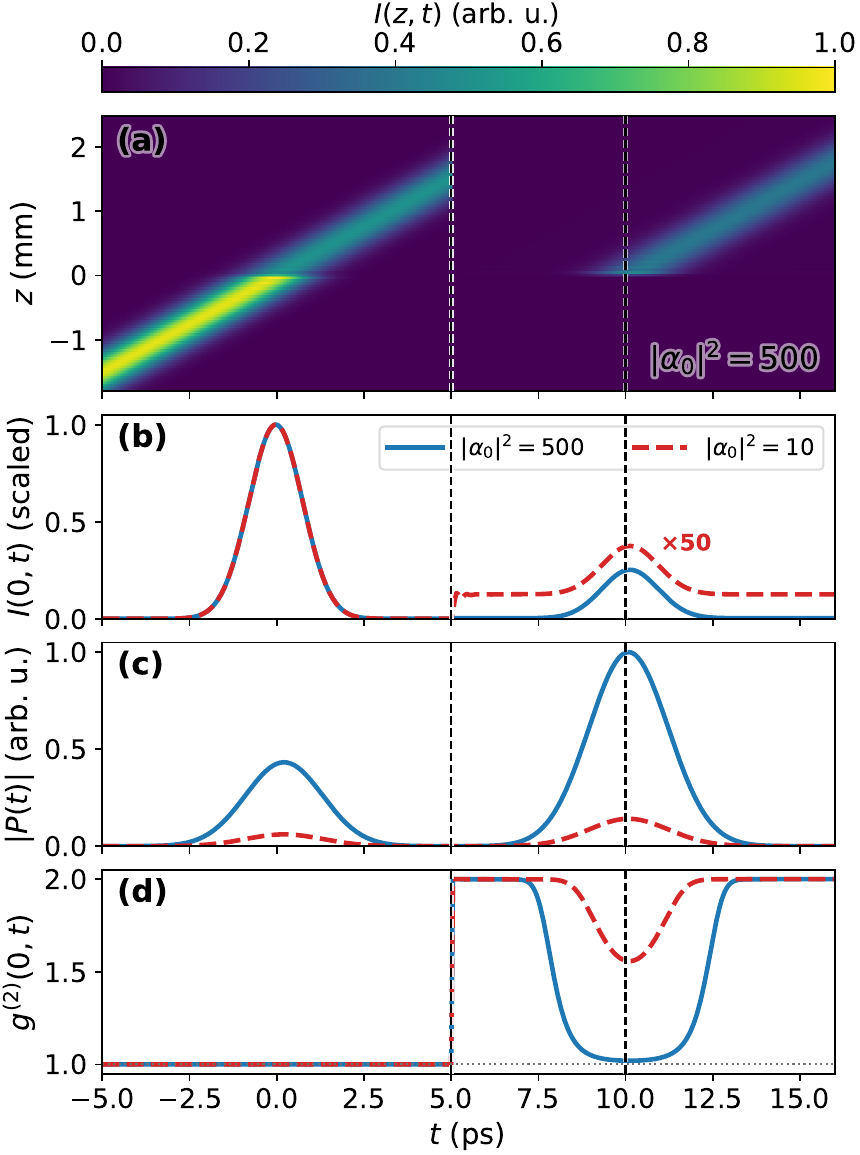}
    \caption{
        Transfer-function calculation of the photon echo. (a) Normally ordered intensity $I(z,t)$ for $|\alpha_0|^2=500$. (b) $I(z=0,t)$ for
        $|\alpha_0|^2=500$ and $|\alpha_0|^2=10$, where the latter is individually scaled by a factor of $50$. (c) Macroscopic polarization $|P(t)|$ for $|\alpha_0|^2=500$ and $|\alpha_0|^2=10$.
        (d) Zero-delay second-order correlation function $g^{(2)}(0,t)$ for $|\alpha_0|^2=500$ and $|\alpha_0|^2=10$. All plots have vertical dashed lines that indicate $\tau_{12} = 5$~ps and $2\tau_{12} = 10$~ps
    }
    \label{fig:echo_transfer_model}
\end{figure}

Figure~\ref{fig:echo_transfer_model}(a) shows the calculated field intensity $I(z,t)$ for $|\alpha_0|^2=500$. The electric field propagates from $z_1 = c t_0<0$ towards the sample at $z=0$. During the interaction with the sample, its amplitude is reduced due to absorption, after which the transmitted field continues to propagate. The excitation at \(t=0\) generates a macroscopic polarization $P(t)$, since the microscopic coherences are created in phase. For $0<t<\tau_{12}$, the microscopic polarizations dephase which leads to a rapid decay of the macroscopic polarization. Due to the action of the $\pi$-pulse, the microscopic polarizations start to rephase at $t=\tau_{12}$. At the rephasing time $t_{\rm echo} = 2\tau_{12}$, all microscopic polarizations are in phase, and, therefore, a macroscopic polarization $P(t)$ is generated. Due to the light-matter coupling, $P(t)$ leads to the emission of a photon echo at $z=0$, from where the echo field propagates. 

Figure~\ref{fig:echo_transfer_model}(b) compares the intensity at the sample position for excitations with $|\alpha_0|^2=500$ and $|\alpha_0|^2=10$, subsequently referred to as strong and weak field, respectively. For $t>\tau_{12}$ the incoherent contributions always generate a noisy background of the intensity, which is clearly visible for the weak field. For the strong field, however, the photon echo is much more pronounced, so that the incoherent contributions are negligible in comparison. Figure~\ref{fig:echo_transfer_model}(c) shows the dynamics of the macroscopic polarization $P(t)$ for $|\alpha_0|^2=500$ and $|\alpha_0|^2=10$, which behaves according to the explanation provided for Fig.~\ref{fig:echo_transfer_model}(a).

Figure~\ref{fig:echo_transfer_model}(d) shows the corresponding second-order correlation function $g^{(2)}(0,t)$. After the $\pi$-pulse, the performed polarization conjugation results in spontaneous emission of photons due to quantum fluctuations \cite{Yeh1992,Schirmer1997} leading to the thermal behavior of $g^{(2)}(0,t)$, which , however, changes in time due to the increase of the coherent contribution $n_{\rm coh}$ that follows the photon echo amplitude.
Hence, $n_{\rm coh}(\tau_{12}+dt) \ll n_{\rm inc}(\tau_{12}+dt)$ for a small but finite $dt$, leading to a value of $g^{(2)}(0,t)\approx2$. During the formation of the photon echo, this value decreases and approaches \(1\) near the echo time \(2\tau_{12}=10~\mathrm{ps}\) for the strong field, since the coherent contribution that follows the echo amplitude exceeds the incoherent ones, i.e. $n_{\rm coh}(2\tau_{12}) \gg n_{\rm inc}(2\tau_{12})$. For the weak field, a decrease is also observed, but $g^{(2)}(0,t)$ does not come very close to $1$, since the coherent photon echo is not significantly stronger than the incoherent contributions. 

The strong field simulations Fig.~\ref{fig:echo_transfer_model} qualitatively correspond to the situation in the experiment and are in agreement with the respective results. We also verified that the echo does not rely on including population inversion in the modeling. Indeed, if the rephasing pulse is considered as a weak pulse that induces the conjugation of the microscopic coherences while keeping $s=-1$ for $t\ge\tau_{12}$, a photon echo with similar $g^{(2)}(0,t)$ dynamics is still obtained, while the incoherent contribution $n_{\rm inc}$ is somewhat reduced. This is consistent with the experimental results for $g^{(2)}(0)$ presented in Tab.~\ref{tab:g2} where pulse areas $\Theta_2$ different from $\pi$ lead to $g^{(2)}(0)\approx 1$ as well.

In summary, the present free-space multimode treatment, including propagation of the quantum-optical field and an idealized $\pi$-pulse rephasing step, demonstrates that photon echoes arising from excitation with coherent light result in $g^{(2)}(0,2\tau_{12}) \approx 1$, in agreement with the experiment.

\begin{table}
\caption{Simulation parameters used in Fig.~\ref{fig:echo_transfer_model}.}
\label{tab:model_parameters}
\begin{ruledtabular}
\begin{tabular}{ll}
Quantity & Value \\
\hline
\(n_\Delta\) (grid points) & \(500\) \\
\(n_\omega\) (grid points) & \(500\) \\
\(\hbar \Delta_{\min},\hbar \Delta_{\max}\) & \(-15,\,15~\mathrm{meV}\) \\
\(\hbar \omega_{\min},\hbar \omega_{\max}\) & \(-15,\,15~\mathrm{meV}\) \\
\(\hbar \mathrm{FWHM}_\Delta\) & \(5~\mathrm{meV}\) \\
\(\hbar \mathrm{FWHM}_\omega\) & \(1~\mathrm{meV}\)\\
\(N_{\mathrm{TLS}}\) & \(2\times10^{6}\) \\
\(\hbar g\) & \(1.7\times10^{-4}~\mathrm{meV}\) \\
\(\tau_{12}\) & \(5~\mathrm{ps}\) \\
\(t_0\) &  \(-5~\mathrm{ps}\) \\
\(|\alpha_0|^2\) & \(500,\;10\)
\end{tabular}
\end{ruledtabular}
\end{table}

\subsection{Discussion}
\label{sec:Disc}

The observation of pronounced Rabi oscillations in Sec.~\ref{sec:Rabi} confirms that coherent excitation and rephasing of the exciton ensemble are working properly. However, the PE signal level obtained in the current experiments appears to be surprisingly low. Taking into account the considerable absorption in the sample ($\alpha L \approx 1.2$ in Fig.~\ref{fig:PL}) and the optimal rephasing condition for the second pulse with an area $\pi$, i.e., when all NCs excited by the first pulse contribute to the two-pulse PE signal, one would expect nearly unity efficiency~\cite{Ruggiero2009}. In this case, according to estimates based on the NC density and the spectral width of the laser, the number of resonantly excited NCs corresponds to  $N_{\rm X} \approx 5\times 10^6$,  implying that the measured signals are at least five orders of magnitude smaller than expected.

We should take into account several factors that reduce the PE efficiency: 
(i) a cross-polarized configuration with an ensemble of randomly oriented NCs yields only $(1/5)^2$ of the total PE intensity, as shown in Ref.~\onlinecite{Trifonov2025arXiv}; 
(ii) imperfect phase matching reduces the signal by a factor of 0.6 (see Sec.~\ref{sec:Setup}); 
(iii) loss of coherence, $\exp(-4\tau_{12}/T_2) = 0.2$, calculated for $\tau_{12} = 73.4$~ps and $T_2 = 180$~ps; 
(iv) optical losses in the detection path correspond to a factor of 0.3; and 
(v) the quantum efficiency is only 0.05 in these particular NCs, further reducing the PE signal~\cite{Kolobkova2021}. 
Taking all these factors into account, we estimate the maximum number of detected photons to be around 300, which is only about one order of magnitude larger than the maximum signal of 30 observed for the $\pi/2$--$\pi$ sequence. Such a difference is acceptable, considering that the exact NC density in the sample is not precisely known. 

Note that the absorption spectrum in Fig.~\ref{fig:PL} does not accurately reflect the number of NCs for which excitons can be treated as two-level systems. For instance, the $\pi/2$ pulse contains approximately $10^{8}$ photons while the number of NCs involved in our measurement is about 20 times smaller. In this case, most absorption occurs in a much larger number of NCs that are excited quasi-resonantly via phonon-assisted transitions, as recently discussed in Ref.~\onlinecite{Trifonov2025arXiv}. In our experiment, only the zero-phonon transitions contribute to the coherent signal. This interpretation is further supported by the observation of strong bleaching and negligible stimulated emission in pump-probe measurements, where no Rabi oscillations are detected in the transmission (see Appendix~\ref{sec:A:bleaching}). These results suggest that only a small fraction of NCs contributes to the PE emission, leading to an overall low efficiency and preventing the detection of other signals, such as incoherent  spontaneous emission with different statistical properties as predicted by theory. To overcome this limitation, further optimization of NC growth toward more homogeneous ensembles is required. Additionally, resonant excitation of trion states could help to suppress fine-structure effects. 

It is worth mentioning that photon echo formation occurs as a result of interference at very low signal levels, with only a few photons per PE pulse. In this regime, we begin to approach the quantum limit without significant noise from spontaneous emission. This feature is inherent to homodyne detection with picosecond time resolution. It allows for filtering of the signal at the appropriate delay, which is two to three orders of magnitude shorter than the exciton lifetime of approximately 1~ns in the studied NCs~\cite{Meliakov2026}.

\section{Conclusions}
\label{sec:Conc}

We have studied the photon statistics of PE signals at cryogenic temperature in an ensemble of CsPbI$_3$ perovskite nanocrystals. We demonstrate coherent control of excitons by detecting pronounced Rabi oscillations of the photon echo for pulse areas up to $2\pi$ under selective resonant excitation of zero-phonon excitons using spectrally narrow picosecond pulses. The damping of the Rabi oscillations is attributed to spatial inhomogeneity of the excitation (transverse beam profile and the finite optical thickness of the sample) as well as to excitation-induced dephasing. The latter is evidenced by the dependence of coherence time on the excitation strength, where $T_2$ decreases non-monotonically from 210 to 100~ps. The low efficiency, due to the small number of nanocrystals involved in the photon echo sequence, results in a low number of photons per pulse despite the significant absorption in the sample. Numerical simulations demonstrate the formation of photon echoes in a quantized free-space multimode description. 

Using an optical homodyne detection scheme, we analyze the photon statistics of the resulting photon echo signal via the second-order correlation function and the characteristic function. Independent of area of either pulse, we consistently measure $g^{(2)}(0)\approx 1$, and the characteristic function remains within the classical regime. These results confirm the classical nature of the photon echo as a coherent emission from an ensemble of emitters under excitation with classical laser pulses. We also note the importance of homodyne detection for reducing noise from spontaneous emission in detection. Future work will focus on applying the presented technique to study other semiconductor nanostructures, as well as different excitonic complexes with higher quantum efficiency and weaker decoherence in ensembles, e.g., trions in lead-bromide perovskite nanocrystals or self-assembled InGaAs quantum dots.

We have numerically analyzed the generation of photon echoes by considering the interaction of multi-mode quantum light with an inhomogeneous distribution of two-level systems. By computing coupled transfer functions for the evoloution of the light field and the material excitations, the dynamics of the respective operators is obtained which allows us to evaluate the time-dependence of several quantities like the macroscopic polarization, the field intensity, and g$^{(2)}$. In agreement with experiment, this approach shows that at the echo time we have essentially g$^{(2)}$(0)=1 for a sufficiently strong coherent excitation.

\begin{acknowledgments}
We acknowledge financial support from the Deutsche Forschungsgemeinschaft (DFG) through the Collaborative Research Centre TRR 142 (project number 231447078, projects A02, A04 and C10).
\end{acknowledgments}

\appendix

\section{Excitation induced dephasing}
\label{sec:A:EID}

\begin{figure}
\includegraphics[width=0.75\linewidth]{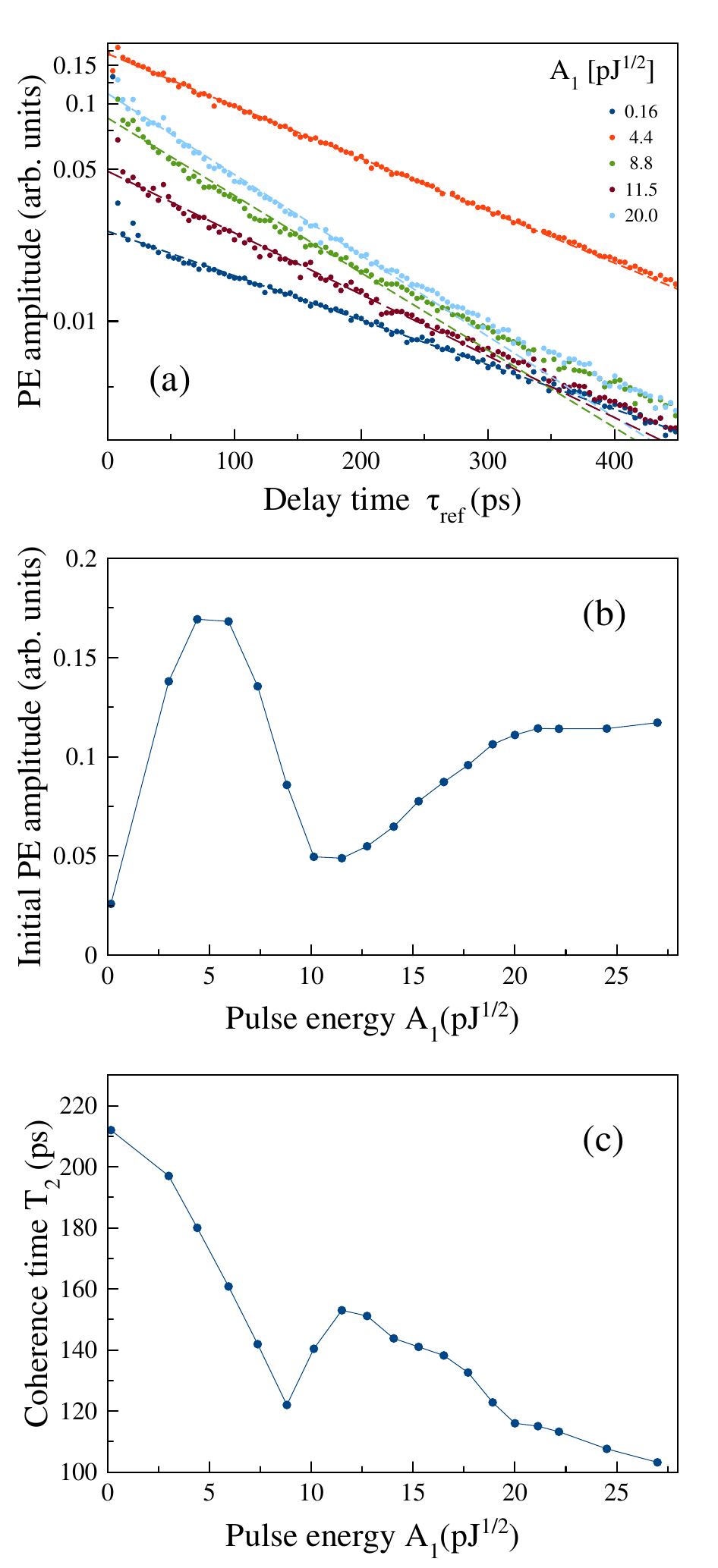}
\caption{(a) Set of PE decays for different $A_1$ and $A_2=9$~pJ$^{1/2}$ ($\Theta_2 \approx \pi$). Dashed lines are fits with exponential decay $P_0\exp(-\tau_{\rm ref}/T_2)$. (b) Initial amplitude $P_0$ as function of $A_1$. (c) Dependence of the exciton coherence time $T_2$ on $A_1$.}
\label{fig:A_EID}
\end{figure}

Here, we discuss the modification of the photon echo decay for different intensities of the first excitation pulse. The data are summarized in Fig.~\ref{fig:A_EID}. For large excitation intensities ($\Theta_1 \gtrsim \pi$), we observe the following features. First, the decay of the PE with $\tau_{\rm ref}=2\tau_{12}$ does not follow a single exponential. At shorter delays, the decay is faster (see Fig.~\ref{fig:A_EID}(a)), indicating that a larger exciton population in the NC leads to shorter coherence time. Second, the dependence of $T_2$ is non-monotonic and shows a dip at $\Theta_1=\pi$ (compare Figs.~\ref{fig:A_EID}(b) and \ref{fig:A_EID}(c)), corresponding to the largest exciton population in the resonantly addressed NCs. 
Both features highlight the role of excitation-induced dephasing, which increases with increasing exciton population.

\section{Bleaching vs amplification }
\label{sec:A:bleaching}

\begin{figure}
\includegraphics[width=\linewidth]{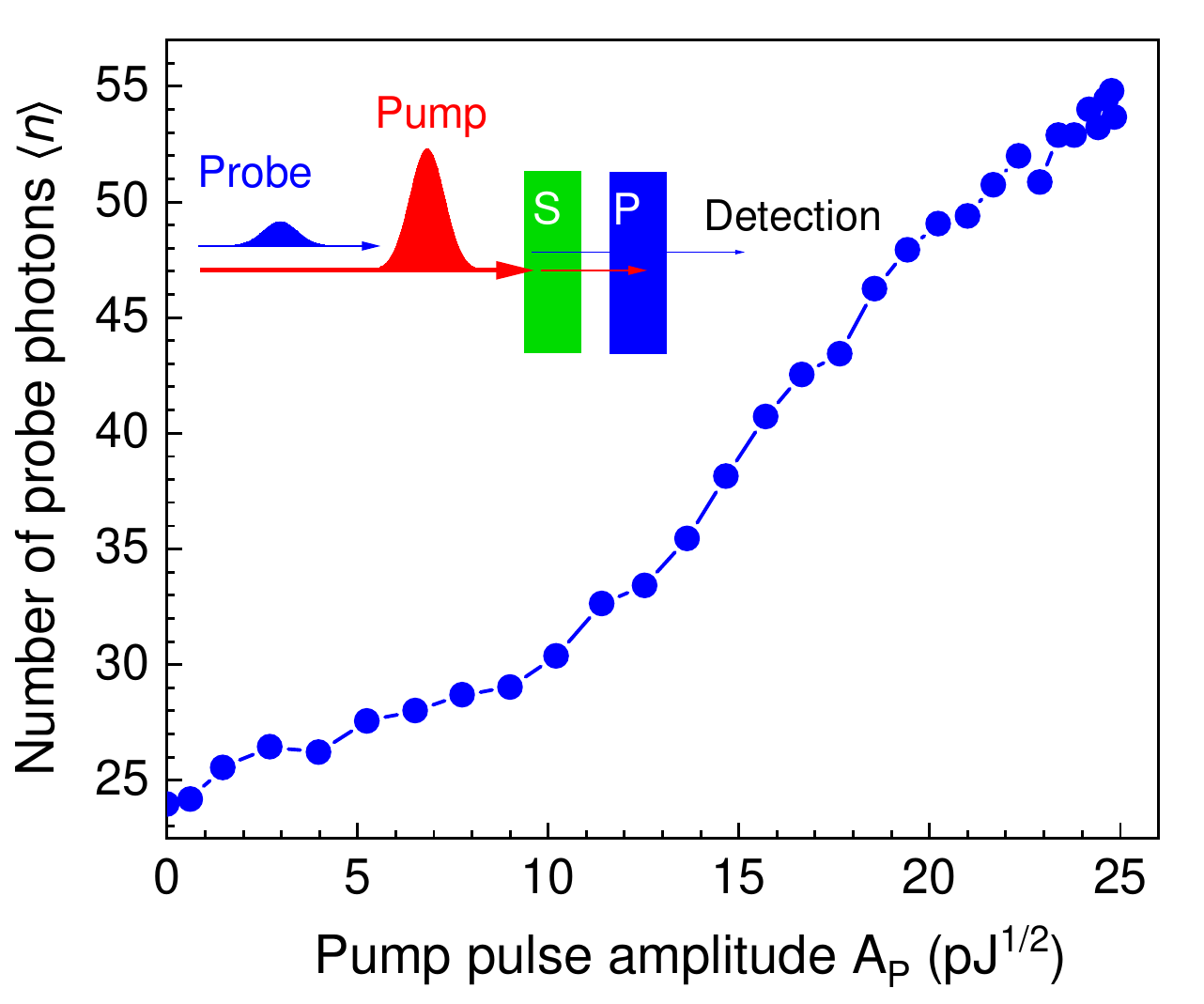}
\caption{Transmitted probe intensity as a function of pump pulse amplitude $A_{\rm P}$. Inset shows the scheme of experiment. S - sample, P - polarizer.}
\label{fig:A_bleach}
\end{figure}

We use a colinear pump-probe scheme to evaluate the bleaching and amplification effects (see inset in Fig.~\ref{fig:A_bleach}). The sample is excited by two cross-polarized laser pulses separated by a delay time of 32~ps. The intensity of the second (probe) pulse is strongly attenuated to about 23 photons per pulse. Its transmitted intensity is recorded as a function of the amplitude of the first (pump) pulse $A_{\rm P}$, which is varied from 0 to 25 pJ$^{1/2}$ (corresponding to the pulse area $\Theta_{\rm P} \approx 0-2.8 \pi$). A polarizer is introduced in the detection path to selectively transmit the probe pulse while strongly suppressing scattered light from the pump pulse.

The dependence of the transmitted intensity of the second pulse on the number of photons is presented in Fig.~\ref{fig:A_bleach}. The signal monotonically increases from 23 to 55 when the pulse area continuously varies from 0 to $2.8\pi$. As seen from Fig.~\ref{fig:A_bleach}, we do not observe Rabi oscillation in the transmitted intensity. If, for $\Theta_{\rm P} = \pi$, all nanocrystals with excitons resonant with the pump photon energy were populated, one would expect a peak in transmission at the pump pulse energy $A_{\rm P}=9$~pJ$^{1/2}$ due to stimulated emission initiated by the probe pulse. This peak would then be followed by a decrease for larger pump pulse areas.
Instead, the transmission monotonically increases by about a factor of 2, i.e., it does not exceed the overall absorption of 70\% which is shown in Fig.~\ref{fig:PL}. This points to the fact that we mainly observe the bleaching of absorption rather than amplification of the second pulse. 

Strong bleaching and negligible amplification can be explained as follows. The number of excitons that contribute to non-linear coherent signals such as PE, i.e., can be described as two-level systems, is significantly smaller than the total number of NCs excited by the laser pulses. Most excitons in NCs undergo fast energy relaxation into the lower-energy states within the first 10~ps~\cite{Trifonov2025arXiv}. The absorption of such NCs is bleached, but they do not contribute to stimulated emission initiated by the second pulse because these excitons are no longer resonant with it. Such NCs also do not contribute to the two-pulse PE signal for $\tau_{12}>10$~ps. As a result, the overall absorption is reduced, but the number of resonant excitons remains too small to produce amplification. This explains the absence of amplified spontaneous emission and the weak photon echo signals observed in the present study, even though the regime of Rabi flops is well established.

\section{Transfer-function method}
\label{app:transfer_functions}

Here we summarize the transfer-function method used in the numerical
simulations. The time evolution is split into two stages. Stage \(j=1\)
describes the evolution before the \(\pi\)-pulse and uses \(s_1=-1\). Stage
\(j=2\) describes the evolution after the \(\pi\)-pulse and uses \(s_2=+1\).
For each stage, the transfer functions are defined by
\begin{align}
\hat{\sigma}_-^{(j,\mathrm{out})}(\Delta,t)
&=
\int d\Delta'\,
K_{MM}^{(j)}(\Delta,\Delta';t)\,
\hat{\sigma}_-^{(j,\mathrm{in})}(\Delta',t_{\rm init})
\nonumber\\
&\quad+
\int d\omega'\,
K_{MF}^{(j)}(\Delta,\omega';t)\,
\hat a^{(j,\mathrm{in})}(\omega',t_{\rm init}),
\label{eq:stagej_sigma_function}
\\
\hat a^{(j,\mathrm{out})}(\omega,t)
&=
\int d\omega'\,
K_{FF}^{(j)}(\omega,\omega';t)\,
\hat a^{(j,\mathrm{in})}(\omega',t_{\rm init})
\nonumber\\
&\quad+
\int d\Delta'\,
K_{FM}^{(j)}(\omega,\Delta';t)\,
\hat{\sigma}_-^{(j,\mathrm{in})}(\Delta',t_{\rm init}).
\label{eq:stagej_a_function}
\end{align}
Here \(t_{\rm init}=t_0\) for \(j=1\) and
\(t_{\rm init}=\tau_{12}\) for \(j=2\).
The equations for the transfer functions follow as
\begin{align}
\dot{ K}_{MM}^{(j)}(\Delta,\Delta';t)
&=
-i\Delta\, K_{MM}^{(j)}(\Delta,\Delta';t)
\nonumber\\
&\quad
+i s_j g\int d\omega\,
K_{FM}^{(j)}(\omega,\Delta';t),
\label{eq:tf_mm}
\\[0.4em]
\dot K_{MF}^{(j)}(\Delta,\omega';t)
&=
-i\Delta\,K_{MF}^{(j)}(\Delta,\omega';t)
\nonumber\\
&\quad
+i s_j g\int d\omega\,
K_{FF}^{(j)}(\omega,\omega';t),
\label{eq:tf_mf}
\\[0.4em]
\dot{ K}_{FF}^{(j)}(\omega,\omega';t)
&=
-i\omega\, K_{FF}^{(j)}(\omega,\omega';t)
\nonumber\\
&\quad
-i g^*\int d\Delta\,
f(\Delta)
K_{MF}^{(j)}(\Delta,\omega';t),
\label{eq:tf_ff}
\\[0.4em]
\dot K_{FM}^{(j)}(\omega,\Delta';t)
&=
-i\omega\,K_{FM}^{(j)}(\omega,\Delta';t)
\nonumber\\
&\quad
-i g^*\int d\Delta\,
f(\Delta)
K_{MM}^{(j)}(\Delta,\Delta';t).
\label{eq:tf_fm}
\end{align}

At the beginning of each stage, the initial conditions are
\begin{align}
K_{MM}^{(j)}(\Delta,\Delta';t_{\rm init})
&=
\delta(\Delta-\Delta'),
\\
K_{FF}^{(j)}(\omega,\omega';t_{\rm init})
&=
\delta(\omega-\omega'),
\\
K_{MF}^{(j)}(\Delta,\omega';t_{\rm init})
&=
0,
\\
K_{FM}^{(j)}(\omega,\Delta';t_{\rm init})
&=
0.
\end{align}

The two stages are connected by the ideal \(\pi\)-pulse at
\(t=\tau_{12}\). The pulse leaves the field unchanged and conjugates the coherence operator,
\begin{align}
\hat a^{(2,\mathrm{in})}(\omega,\tau^+_{12})
&=
\hat a^{(1,\mathrm{out})}(\omega,\tau^-_{12}),
\\
\hat{\sigma}_-^{(2,\mathrm{in})}(\Delta,\tau^+_{12})
&=
\left[
\hat{\sigma}_-^{(1,\mathrm{out})}(\Delta,\tau^-_{12})
\right]^\dagger .
\end{align}
For the photon-echo calculation we separate the emitted echo mode from the
incident field mode. The echo mode is initially in the vacuum state. With this
choice, the photon annihilation operator after the second stage can be written as
\begin{align}
\hat b(\omega,t)
&=
\int d\omega'\,
K_{FF}^{(2)}(\omega,\omega';t)
\hat b^{\mathrm{in}}(\omega',\tau_{12})
\nonumber\\
&\quad+
\int d\omega'\,
B(\omega,\omega';t)\hat a^\dagger(\omega',t_0)
\nonumber\\
&\quad+
\int d\Delta'\,
Y(\omega,\Delta';t)
\hat{\sigma}_-^\dagger(\Delta',t_0).
\label{eq:app_b_operator}
\end{align}
Here the first term is connected with the vacuum contribution. The functions
\(B\) and \(Y\) are obtained by combining the first-stage matter functions with
the second-stage field--matter function,
\begin{align}
B(\omega,\omega';t)
&=
\int d\Delta\,
K_{FM}^{(2)}(\omega,\Delta;t)
\left[
K_{MF}^{(1)}(\Delta,\omega';\tau_{12})
\right]^*,
\label{eq:app_B_function}
\\
Y(\omega,\Delta';t)
&=
\int d\Delta\,
K_{FM}^{(2)}(\omega,\Delta;t)
\left[
K_{MM}^{(1)}(\Delta,\Delta';\tau_{12})
\right]^* .
\label{eq:app_Y_function}
\end{align}
All integrals are evaluated using the composite trapezoidal rule on the grids listed in
Table~\ref{tab:model_parameters}.
In the numerical representation, the transfer-function initial conditions are
identity matrices on these grids.

We consider the second-order correlation function for zero delay at the sample position, which is defined as
\begin{align}
    g^{(2)}(0,t) &= \frac{\langle \hat E^{(-)}(0,t) \hat E^{(-)}(0,t) \hat E^{(+)}(0,t) \hat E^{(+)}(0,t) \rangle}{\langle \hat E^{(-)}(0,t) \hat E^{(+)}(0,t) \rangle\langle \hat E^{(-)}(0,t) \hat E^{(+)}(0,t) \rangle}.\label{eq:g2_definition}
\end{align}
We assume $\mathcal{E}(\omega)$ to be constant in the region of interest, after which Eq.~\eqref{eq:g2_definition} simplifies to Eq.~\eqref{eq:model_g2_main} with the following abbreviations

\begin{align}
n_{\mathrm{coh}}(t)
&=
|\bar{\alpha}(t)|^2,
\\
n_{\mathrm{inc}}(t)
&=
\bar B(t)+\bar Y(t).
\end{align}

\begin{align}
\bar{\alpha}(t)
&=
\int d\omega\,d\omega'\,
B^*(\omega,\omega';t)\alpha(\omega'),
\\
\bar B(t)
&=
\int d\omega_1d\omega_2d\omega'\,
B^*(\omega_1,\omega';t)B(\omega_2,\omega';t),
\\
\bar Y(t)
&=
\int d\omega_1d\omega_2d\Delta'\,
\frac{
Y^*(\omega_1,\Delta';t)Y(\omega_2,\Delta';t)
}{f(\Delta')} .
\end{align}

\bibliographystyle{apsrev4-2}

\end{document}